\documentclass{emulateapj}

\newcommand{\mum}{\ifmmode{\rm \mu m}\else{$\mu$m}\fi}
\newcommand{\iras}{{\it IRAS}}
\newcommand{\iso}{{\it ISO}}
\newcommand{\msx}{{\it MSX}}

\begin{document}

\title{Mid-infrared spectroscopy of carbon stars in the Small Magellanic Cloud}

\author{
G.~C.~Sloan\altaffilmark{1},
K.~E.~Kraemer\altaffilmark{2},
M.~Matsuura\altaffilmark{3,4},
P.~R.~Wood\altaffilmark{5},
S.~D.~Price\altaffilmark{2},
M.~P.~Egan\altaffilmark{6}
}
\altaffiltext{1}{Cornell University, Astronomy Department,
  Ithaca, NY 14853-6801, sloan@isc.astro.cornell.edu}
\altaffiltext{2}{Air Force Research Laboratory, Space Vehicles
   Directorate, 29 Randolph Road, Hanscom AFB, MA 01731,
   kathleen.kraemer@hanscom.af.mil, steve.price@hanscom.af.mil}
\altaffiltext{3}{School of Physics and Astronomy, University of Manchester, 
   P.O. Box 88, Sackville Street, Manchester M60 1QD, United Kingdom}
\altaffiltext{4}{Department of Physics and Astronomy, Queen's University
   of Belfast, Belfast BT7 1NN, Northern Ireland, United Kingdom,
   m.matsuura@qub.ac.uk}
\altaffiltext{5}{Research School of Astronomy and Astrophysics,
  Australian National University, Cotter Road, Weston Creek ACT 2611,
  Australia, wood@mso.anu.edu.au}
\altaffiltext{6}{Air Force Research Laboratory, Space Vehicles
  Directorate, 1851 S. Bell St., CM3, Suite 700, Arlington, VA
  22202, michaelegan.cox.net}


\slugcomment{PREPRINT}

\begin{abstract}

We have observed a sample of 36 objects in the Small Magellanic 
Cloud (SMC) with the Infrared Spectrometer on the {\it Spitzer 
Space Telescope}.  Nineteen of these sources are carbon stars.
An examination of the near- and mid-infrared photometry shows
that the carbon-rich and oxygen-rich dust sources follow two 
easily separated sequences.  A comparison of the spectra of the 
19 carbon stars in the SMC to spectra from the {\it Infrared 
Space Observatory} (\iso) of carbon stars in the Galaxy reveals 
significant differences.  The absorption bands at 7.5~\mum\ and 
13.7~\mum\ due to C$_2$H$_2$ are stronger in the SMC sample, and 
the SiC dust emission feature at 11.3~\mum\ is weaker.  Our 
measurements of the MgS dust emission feature at 26--30~\mum\ 
are less conclusive, but this feature appears to be weaker in 
the SMC sample as well.  All of these results are consistent 
with the lower metallicity in the SMC.  The lower abundance
of SiC grains in the SMC may result in less efficient 
carbon-rich dust production, which could explain the excess
C$_2$H$_2$ gas seen in the spectra.  The sources in the SMC 
with the strongest SiC dust emission tend to have redder 
infrared colors than the other sources in the sample, which 
implies more amorphous carbon, and they also tend to show 
stronger MgS dust emission.  The weakest SiC emission features 
tend to be shifted to the blue; these spectra may arise from
low-density shells with large SiC grains.

\end{abstract}

\keywords{ circumstellar matter --- infrared:  stars --- 
stars:  carbon --- Magellanic Clouds}

\section{Introduction} 

The formation of CO in the cool atmospheres of stars on the
asymptotic giant branch (AGB) forces the chemistry of the 
dust produced by mass-losing stars to be either oxygen-rich
or carbon-rich, depending on whether the C/O ratio in these
stars is less than or more than one.  In the Galaxy, AGB
stars are the primary source of material injected into the
interstellar medium (ISM) \citep{geh89}, and they are
predominantly oxygen-rich \citep[see the statistics of the
characterizations of spectra from the \iras\ Low-Resolution
Spectrometer by the][]{lrs86}.  Consequently, carbon-rich dust
species are a small fraction of the material injected into the
galactic ISM compared to silicates and related oxygen-rich 
dust species.

The Large Magellanic Cloud (LMC) has a higher fraction of 
carbon stars than the Galaxy \citep{bbm78}, and the Small 
Magellanic Cloud (SMC) has a higher fraction still 
\citep{bbm80}.  \cite{rv81} showed that reducing the initial 
metallicity of a star increases the likelihood that it will 
become a carbon star on the AGB.  More metal-poor stars have 
less oxygen in their atmospheres initially, so once they 
begin to dredge up carbon-rich material from the interior, 
fewer dredge-ups are needed to raise the C/O ratio past unity
\citep{lw04}.  As a result, the mass limit above which stars 
will evolve into carbon stars decreases as metallicity 
decreases.  In the models with Z=0.02 examined by 
\cite{rv81}, the limit is $\sim$1.8~M$_\sun$ (depending on 
other assumptions), but for Z=0.004, the limit is 
$\sim$1.3~$M_\sun$.


Thus, in low-metallicity galaxies, more of the dust injected
into the ISM from AGB stars will be carbon-rich.  The {\it
Spitzer Space Telescope} \citep{wer04} can observe
fainter, more distant, and more metal-poor galaxies than
previous infrared telescopes, which makes it essential to
better understand the nature of the dust produced in these
systems.  The sensitivity of the Infrared Spectrograph 
\citep[IRS,][]{hou04} on {\it Spitzer} makes it possible to 
obtain spectra of a large sample of individual dust shells 
around evolved stars in the Magellanic Clouds.  The metallicity
of the younger population in the SMC is $\sim$0.2 Solar 
\citep{rb89,hi97,lu98}.  The sources which populate the AGB 
are older and may even be more metal poor.  Thus the SMC is an
ideal laboratory for studying how metallicity affects
the nature of carbon-rich dust produced on the AGB.  


We have used the IRS to observe a sample of 36 infrared 
sources in the SMC, most of which are evolved stars.  Nineteen
of these sources are carbon-rich AGB sources.  In this paper, 
we will compare their spectroscopic properties to the 
Galactic sample of carbon stars, as observed by the 
Short-Wavelength Spectrometer (SWS) aboard \iso.  The spectra 
from carbon stars in the wavelength range covered by both
instruments include both emission features from dust and
absorption features from molecular gas.

\cite{gil69} predicted that the dominant condensates in the
outflows of carbon stars would be carbon (later specified 
as graphite or amorphous carbon) and silicon carbide. 
\cite{hac72} first observed an emission feature at 11.3~\mum\
in the carbon stars V Hya and CIT 6.  \cite{tc74} confirmed
the presence of this feature in spectra of CIT 6 and
IRC +10216, and they identified its carrier as SiC dust.
Amorphous carbon dominates the dust produced by heavily 
reddened carbon stars, as shown by models of the spectral 
energy distribution of the extreme carbon star IRC +10216 
\citep[e.g.,][]{mr87,gri90,se95}.  

\cite{gm85} identified MgS dust as the carrier of a broad
emission feature first observed spectroscopically in the
26--30~\mum\ range by \cite{fhm81} in the shells around
heavily reddened carbon stars.  SWS observations of 
Galactic carbon stars served as the basis for a thorough 
study by \cite{hwt02}.  They found a wide variation in the 
shape of the feature, which they explained on the basis of 
grain temperature.  Their ability to fit a variety of 
spectral feature shapes using MgS considerably strengthens 
its identification as the carrier of the 26--30~\mum\ 
feature.  
 
In addition to the dust features, the IRS wavelength
range also includes two molecular bands seen in
carbon-rich spectra.  The narrow band at 13.7~\mum\ is
well characterized and arises from acetylene (C$_2$H$_2$),
specifically from the Q branch of the $\nu_5$ band \citep{ato99}.  
The P and R branches produce broad shoulders to this narrow 
feature which typically absorb in the 12.5--15.0~\mum\ 
range.  The identification of the band at 7.5~\mum\ is 
more problematic, as HCN, C$_2$H$_2$, and CS all absorb in 
this spectral region \citep{ato98,jhl00}.  \cite{mat06} 
argue that C$_2$H$_2$ is the dominant absorber, based on 
the similarity of its expected band shape and position to 
the observed band in carbon stars in the LMC.  \cite{mat06}
also show that the HCN band at 14.1~\mum\ seen in Galactic
carbon stars is absent in the LMC sample.  It does not 
appear in our SMC sample, either.

\section{Observations} 

\subsection{The spectral data} 


\begin{deluxetable*}{lllrrrcc} 
\tablenum{1}
\tablecolumns{8}
\tablewidth{0pt}
\small
\tablecaption{Sources and observing information\label{Tbl1}}
\tablehead{
  \colhead{MSX} & \colhead{RA} & \colhead{Dec.} & 
  \multicolumn{2}{c}{2MASS\tablenotemark{a}} &
  \colhead{MSX } & \multicolumn{2}{c}{Integration time (s)} \\
  \colhead{SMC} & \multicolumn{2}{c}{J2000.0} & 
  \colhead{J} & \colhead{K$_s$} & \colhead{A\tablenotemark{b}} & 
  \colhead{SL} & \colhead{LL}
}
\startdata
033 & 00 47 05.52 & $-$73 21 33.0 & 13.445 & 10.320 &  7.320 & 112 & 168 \\
036 & 00 45 53.95 & $-$73 23 41.2 & 15.074 & 11.599 &  7.999 & 240 & 960 \\
044 & 00 43 39.58 & $-$73 14 57.6 & 12.567 & 10.031 &  7.752 & 112 & 240 \\
054 & 00 43 05.90 & $-$73 21 40.6 & 16.597 & 12.573 &  7.794 & 240 & 480 \\
060 & 00 46 40.43 & $-$73 16 47.2 & 15.680 & 11.199 &  6.150 & 112 & 112 \\
062 & 00 42 40.91 & $-$72 57 05.8 & 13.220 & 10.163 &  6.821 & 112 & 112 \\
066 & 00 48 52.51 & $-$73 08 56.8 & 14.699 & 11.115 &  7.184 & 240 & 360 \\
091 & 00 36 56.71 & $-$72 25 17.6 & 13.684 & 10.782 &  7.948 & 112 & 240 \\
093 & 00 59 23.36 & $-$73 56 01.0 & 14.419 & 11.414 &  7.840 & 240 & 960 \\
105 & 00 45 02.15 & $-$72 52 24.3 & 15.305 & 11.192 &  6.813 & 112 & 240 \\
142 & 00 51 40.47 & $-$72 57 29.0 & 12.902 & 10.734 &  8.032 & 112 & 360 \\
159 & 00 54 22.29 & $-$72 43 29.7 & 16.050 & 11.713 &  7.357 & 240 & 360 \\
162 & 00 52 40.18 & $-$72 47 27.7 & 12.482 &  9.846 &  7.375 & 112 & 168 \\
163 & 00 51 00.75 & $-$72 25 18.6 & 14.735 & 10.989 &  7.304 & 240 & 480 \\
198 & 00 57 10.98 & $-$72 31 00.0 & 14.692 & 11.418 &  7.719 & 240 & 960 \\
200 & 00 46 50.79 & $-$71 47 39.3 & 15.120 & 11.460 &  7.800 & 240 & 960 \\
202 & 00 53 10.14 & $-$72 11 54.7 & 11.930 &  9.775 &  7.967 & 112 & 240 \\
209 & 00 56 16.39 & $-$72 16 41.3 & 13.771 & 10.606 &  7.222 & 112 & 168 \\
232 & 01 06 03.30 & $-$72 22 32.3 & 14.302 & 11.318 &  7.317 & 240 & 480 
\enddata
\tablenotetext{a}{From \cite{2mass}.}
\tablenotetext{b}{Magnitude at 8.3~\mum; from \cite{msx03}}
\end{deluxetable*}

The {\it Spitzer Space Telescope} observed all of the sources
in our sample at low resolution (R$\sim$60--120), using the 
standard IRS staring mode.  The integration times for both the 
Short-Low (SL) and Long-Low (LL) modules are split evenly 
between the first- and second-order apertures.  Table 1 
presents the observational details along with photometry at J 
and K$_s$ from 2MASS and \msx\ band A (8.3 \mum).  

The analysis began with the two-dimensional flatfielded
images produced by version S11.0 of the data reduction
pipeline at the {\it Spitzer} Science Center (SSC)\footnote{One
source, MSX SMC 093, was observed later and processed with 
the S12.0 version of the pipeline.}.  We subtracted sky images 
to remove any background emission.  SL images were differenced 
aperture-by-aperture (i.e. SL1$-$SL2 and vice versa), and LL 
images were differenced nod-by-nod.  Before extracting spectra 
from the images, we used the {\it imclean} software package 
distributed by Cornell University\footnote{An updated version 
of this package, now known as {\it irsclean}, is available 
from the SSC website.} to replace bad pixels with values 
determined from neighboring pixels.

To extract spectra from the cleaned and differenced images,
we used the SSC pipeline modules {\it profile}, {\it ridge}
and {\it extract}, which are available through SPICE (the 
Spitzer IRS Custom Extractor, distributed by the SSC).  To 
photometrically calibrate the spectra, we used spectral 
corrections generated from observations of standard stars.  
For SL, we used HR 6348 (K0 III), and for LL, we used HR 6348, 
HD 166780 (K4 III) and HD 173511 (K5 III).  The assumed truth 
spectra of these spectra are modified from spectral templates 
provided by M. Cohen.  A paper explaining this process is in
preparation (Sloan et al.).  We have used the modified 
wavelength calibration of the S11.0 version of the SSC 
pipeline described by \cite{sl05}.

After calibrating the spectra, we co-added them, stitched 
the individual spectral segments together to correct 
discontinuities by making scalar multiplicative corrections,
and trimmed poor data from the ends of the segments.  The 
discontinuities arise primarily from variations in throughput 
caused by pointing variations, since a source will be better 
centered in some apertures than in others.  Typical 
multiplicative corrections varied between 5 and 10\%.  
Segments are normalized upwards to align with the 
best-centered segment.  We have used the bonus-order data in 
SL and LL by combining it with orders 1 and 2 where they 
overlap and are valid.  The error bars are the statistical 
differences between the two nod positions. 

\subsection{The full sample} 

\begin{figure} 
\includegraphics[width=3.5in]{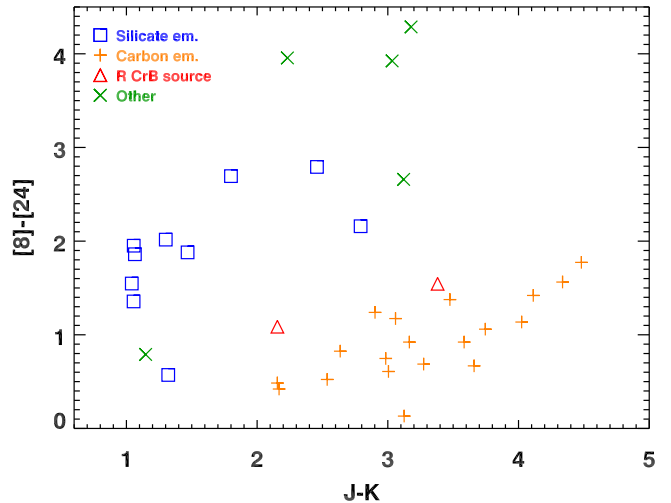}
\caption{A color-color plot of the SMC sample.  Plotting
mid-infared colors vs. near-infrared colors separates the
various dust compositions cleanly.  This example uses
2MASS colors on the horizontal axis and colors derived
from our IRS spectra using the bandpasses for the IRAC 
8~\mum\ and MIPS 24~\mum\ filters on the {\it Spitzer Space
Telescope}.  The carbon-rich dust sources fall along a
clear sequence which is red in J$-$K but blue in [8]$-$[24]
compared to the oxgyen-rich sources.}
\end{figure}

Our full {\it Spitzer}/IRS sample consists of 36 infrared
sources in the SMC.  They were selected on the basis 
of near-infrared colors observed in the 2MASS survey 
\citep{2mass} and mid-infrared measurements from the {\it 
Midcourse Space Experiment} \citep[{\it MSX},][]{msx03}.
The use of the 2MASS survey introduces a selection effect
against heavily enshrouded sources, since reddening from
the dust shell could push the near-infrared colors below
the 2MASS detection limit of K$\sim$16.  The MSX sensitivity
limits the sample to magnitudes at 8~\mum\ brighter than
$\sim$8.5, which selects against sources with little dust.

The first objective of the program was to validate the 
spectroscopic properties predicted from the near-infrared 
and {\it MSX} colors for a sample of sources in the LMC 
\citep{msx01}.  Out of the 36 targets in the SMC, 22 show 
infrared spectral features from carbon-rich dust or gas 
species.  This fraction is higher than expected, primarily 
because the reddest J$-$K sources selected had been
tentatively identified as OH/IR stars.  They
turned out to be embedded carbon stars instead.  As a 
result, our sample of carbon stars covers a wide range of 
infrared colors.


Figure 1 illustrates how a combination of 2MASS and
{\it Spitzer} photometry can distinguish the dust composition 
in evolved stars.  The [8]$-$[24] color is based on synthetic 
photometry of the IRS spectra using the bandpasses for the 
IRAC 8~\mum\ and the MIPS 24~\mum\ filters.  The carbon-rich
dust sources occupy a clear seqence with the [8]$-$[24] color
steadily increasing from $\sim$0 to $\sim$2 as the J$-$K color
increases from $\sim$2 to $\sim$5.  The oxygen-rich sources
occupy a different sequence with J$-$K colors $\sim$1.0--1.5 for
[8]$-$[24] $\lesssim$2 and showing a wide range in J$-$K for 
redder [8]$-$[24] colors.  At no point in Figure 1 do the two 
sequences overlap.  Van Loon et al. (1997) showed that plotting K$-$[12] 
vs. J$-$K produces two separate sequences for the carbon-rich
and oxygen-rich dust sources.  Shifting the longer-wavelength
color further to the infrared, as Figure 1 does here, eliminates
the overlap between the sequences.

Two of the 22 carbon-rich sources exhibit featureless spectra 
resembling $\sim$600--700 K blackbodies.  As described by 
\cite{kra05}, they are R CrB candidates and are thus not 
associated with the AGB.  These sources appear as triangles in 
Figure 1, and they lie slightly above the carbon-rich sequence.  
One of the carbon-rich sources shows PAH emission features 
superimposed on an otherwise smooth continuum peaking in the 
mid-infared.  Kraemer et al. (in preparation) examine this 
source in a separate paper and conclude it has most likely 
evolved off of the AGB.  It is plotted as one of the ``Other'' 
sources in Figure 1.  The remaining 19 sources constitute our 
sample of carbon-stars in the SMC.

\subsection{The carbon stars} 

\begin{figure*} 
\includegraphics[width=6.5in]{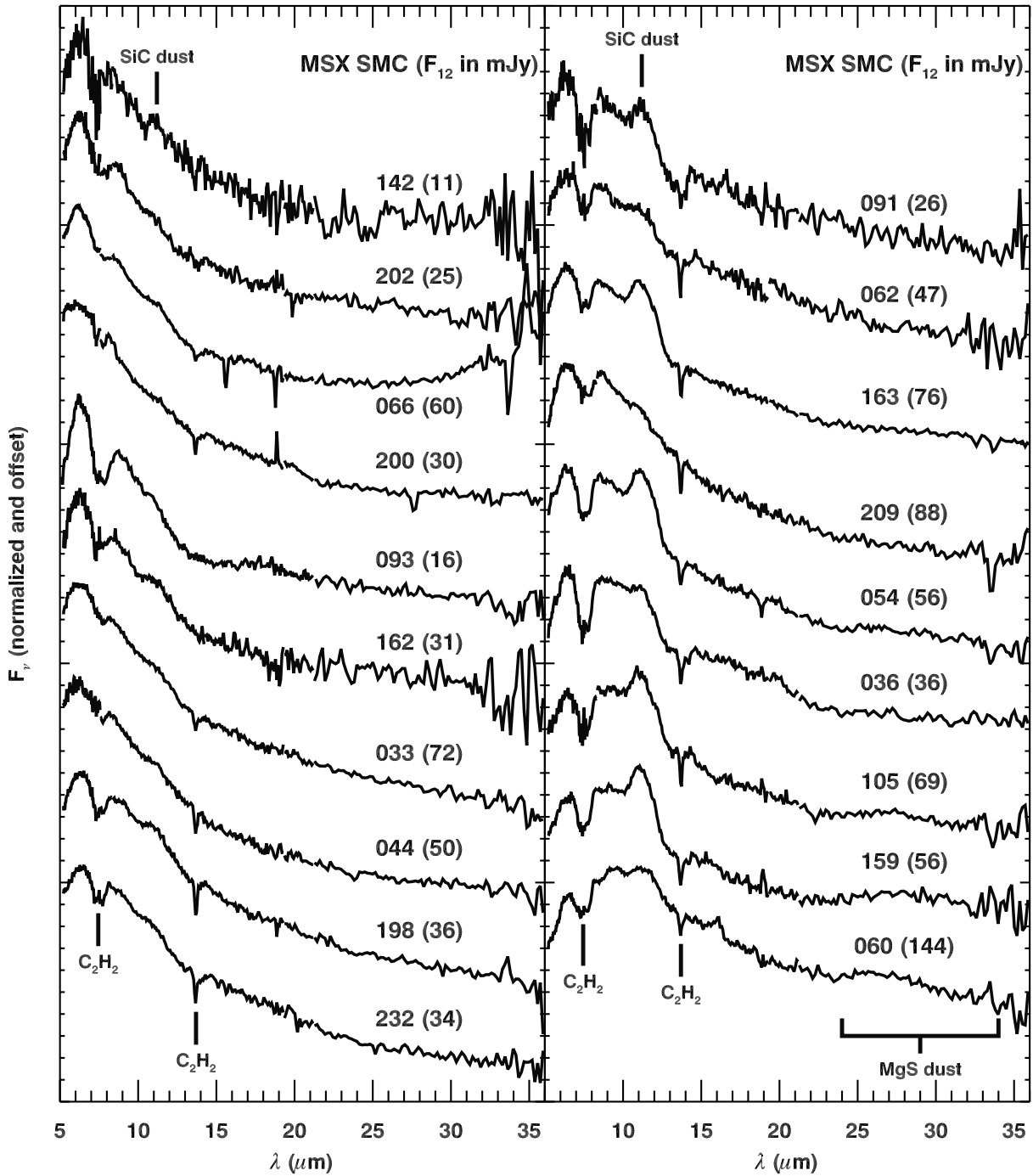}
\caption{Our sample of mid-infrared spectra from carbon stars
in the SMC.  The dominant features are the C$_2$H$_2$
absorption bands at 7.5 and 13.7~\mum, the SiC dust emission
feature at 11.3~\mum, and the MgS dust emission feature in
the 26--30~\mum\ range.  The spectra are ordered by
[6.4]$-$[9.3] color and normalized.  The 12~\mum\ flux in
mJy is included in parentheses after the \msx\ SMC number of
each source.  Some of the spectra show artifacts due to
background emission from forbidden lines that did not fully
cancel out in the differenced images (most notably MSX SMC 
066, with artifacts from [Ne III] at 15.6~\mum\ and [S III] 
at 18.7 and 33.5~\mum).  Other artifacts appear due to 
unflagged misbehaving (i.e. rogue) pixels.}
\end{figure*}

Figure 2 presents the IRS low-resolution spectra of the 19
carbon stars in our sample from the SMC.  All of these
sources exhibit spectral characteristics typical of carbon
stars on the AGB, including molecular absorption bands from 
C$_2$H$_2$ at 7.5 and 13.7~\mum\ and the SiC dust emission 
feature at 11.3~\mum.  A few of the spectra also show a MgS 
dust emission feature peaking in the 26--30~\mum\ range.
Eighteen of the sources were observed in 2004 November; one 
object, MSX SMC 093, was observed in 2005 June.  Figure 2
organizes the spectra by their [6.4]$-$[9.3] color, as 
defined in \S 3.1.

\subsection{Ground-based observations} 

All of the carbon stars in our sample were observed from the
ground quasi-simultaneously with the {\it Spitzer} 
observations.  Near-infrared photometric observations were 
taken with the 2.3-m Telescope at Siding Spring Observatory 
(SSO) in Australia, using the near-infrared imaging system 
CASPIR \citep{caspir} and the filters J (effective wavelength 
1.24~\mum), H (1.68~\mum), K (2.22~\mum), and narrow-band L 
(3.59~\mum).  Standard stars from the lists by \cite{mcg94} 
were used to calibrate the observations.  Standard image 
reduction procedures of bias subtraction, linearization, 
bad pixel replacement, flat fielding and sky subtraction were 
done using {\it IRAF}.  Aperture photometry was performed 
with the {\it IRAF} task {\it QPHOT}.  All objects were 
observed twice and the photometry averaged.  


Table 2 presents the ground-based photometry for our sample.  
The errors given are the worst of the noise estimate from 
{\it QPHOT} or the standard error computed from the two 
observations.  The 18 sources observed in 2004 November by 
{\it Spitzer} were observed four weeks later at SSO.  
MSX SMC 093 was observed by {\it Spitzer} on 2005 June 4.  
The photometric data in Table 2 for this source 
are linearly interpolated from SSO data obtained on May 25 
and July 22.  

\begin{figure} 
\includegraphics[width=3.5in]{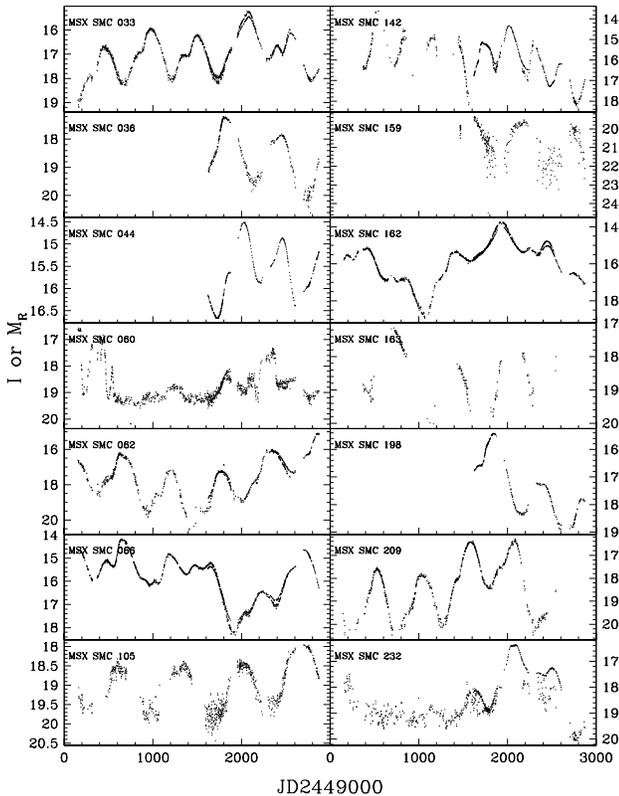}
\caption{Light curves for the {\it Spitzer} sources which 
could be found in the MACHO and/or OGLE databases.  JD2449000 
is Julian Date$-$2449000 (days).  The MACHO data extend over 
the time interval 100 $<$ JD2449000 $<$ 2600, while the OGLE 
data cover the time interval 1600 $<$ JD2449000 $<$ 2900.  
When both MACHO and OGLE data exist for a star, the zero 
point of the MACHO R magnitudes $M_{\rm R}$ has been adjusted 
so that the OGLE I magnitudes and MACHO R magnitudes match on 
average.}
\end{figure}

\begin{deluxetable*}{lcccccr} 
\tablenum{2}
\tablecolumns{7}
\tablewidth{0pt}
\tablefontsize{small} 
\tablecaption{Photometric results\label{Tbl2}}
\tablehead{
  \colhead{MSX} & \colhead{J} & \colhead{H} & \colhead{K} & 
  \colhead{L} & \colhead{period} & \colhead{$L_{bol}$} \\
  \colhead{SMC} & \colhead{(mag)} & \colhead{(mag)} & \colhead{(mag)} & 
  \colhead{(mag)} & \colhead{(days)} & \colhead{(L$_{\sun}$)} \\
}
\startdata
033 & 14.362$\pm$0.022 & 12.404$\pm$0.014 & 10.846$\pm$0.008 &  8.775$\pm$0.042 
    & 530 & 20090 \\
036 & \nodata          & 14.400$\pm$0.024 & 12.711$\pm$0.028 & 10.373$\pm$0.095 
    & 640 &  5480 \\
044 & 15.579$\pm$0.156 & 13.595$\pm$0.025 & 11.806$\pm$0.009 &  9.533$\pm$0.087 
    & 460 & 11780 \\
054 & 16.516$\pm$0.325 & 14.326$\pm$0.044 & 12.247$\pm$0.012 &  9.841$\pm$0.223 
    & \nodata &  8000 \\
060 & 15.011$\pm$0.035 & 12.864$\pm$0.016 & 11.048$\pm$0.007 &  9.119$\pm$0.052 
    & 350 & 17520 \\
062 & 15.343$\pm$0.033 & 13.235$\pm$0.019 & 11.581$\pm$0.009 &  9.501$\pm$0.060 
    & 570 &  9120 \\
066 & 13.305$\pm$0.027 & 11.591$\pm$0.013 & 10.263$\pm$0.007 &  8.574$\pm$0.058 
    & 530 &  6300 \\
091 & 15.311$\pm$0.077 & 13.334$\pm$0.031 & 11.654$\pm$0.011 &  9.917$\pm$0.145 
    & \nodata & 5110 \\
093 & 13.053$\pm$0.022 & 11.516$\pm$0.034 & 10.403$\pm$0.032 &  9.133$\pm$0.080 
    & \nodata & 8210 \\
105 & 16.082$\pm$0.129 & 13.687$\pm$0.024 & 11.783$\pm$0.009 &  9.540$\pm$0.046 
    & 670 &  9120 \\
142 & 14.129$\pm$0.021 & 12.615$\pm$0.036 & 11.404$\pm$0.008 & 10.479$\pm$0.227 
    & 300 &  5180 \\
159 & 17.398$\pm$0.199 & 15.010$\pm$0.078 & 12.656$\pm$0.017 &  9.534$\pm$0.070 
    & 560 &  7470 \\
162 & 13.850$\pm$0.019 & 11.963$\pm$0.025 & 10.575$\pm$0.007 &  8.986$\pm$0.113 
    & 520 & 11480 \\
163 & 15.528$\pm$0.044 & 13.280$\pm$0.040 & 11.526$\pm$0.023 &  9.266$\pm$0.090 
    & 660 & 13000 \\
198 & 14.659$\pm$0.035 & 12.865$\pm$0.016 & 11.442$\pm$0.015 &  9.355$\pm$0.099 
    & 500 &  7810 \\
200 & 16.697$\pm$0.086 & 14.182$\pm$0.020 & 12.207$\pm$0.025 &  9.736$\pm$0.074 
    & \nodata & 9130 \\
202 & 12.974$\pm$0.017 & 11.415$\pm$0.013 & 10.423$\pm$0.011 &  8.998$\pm$0.044 
    & \nodata & 11460 \\
209 & 14.035$\pm$0.025 & 12.147$\pm$0.014 & 10.795$\pm$0.007 &  8.908$\pm$0.042 
    & 520 & 16330 \\
232 & 15.539$\pm$0.085 & 13.726$\pm$0.042 & 12.029$\pm$0.024 &  9.726$\pm$0.286 
    & 460 &  7280 
\enddata
\end{deluxetable*}


Of the 19 carbon stars in our sample, 14  have light 
curves in the databases produced by the MACHO Project 
\citep[Massive Compact Halo Objects;][]{macho} and/or the 
Optical Gravitational Lensing Experiment 
\citep[OGLE;][]{ogle97,ogle05}.  Figure 3 presents these 
lightcurves.  These stars all show evidence of pulsation, 
with a typical period of $\sim$300--700 days, making them 
large-amplitude, long-period variables.

These variables are typical of large-amplitude, dust 
enshrouded stars found in the Magellanic Clouds and the 
Galaxy \citep{whi03,lw04}.  As well as the main pulsation 
period of $\sim$300--700 days, they often show long periods 
of dimming and brightening of many thousands of days (MSX SMC 
060, 062, 142, 162, 232) and double-humped light curves, i.e. 
alternate faint and bright minima (MSX SMC 033, 066, 209).  
\cite{ow05} find this latter effect in their models of 
large-amplitude pulsating giants.  

Our carbon stars are very red and fainter in K than stars of 
similar bolometric luminosity, but with shorter period and lower 
mass-loss rate, and as a result, on the K-$\log P$ diagram, 
they typically lie on a very steep sequence of declining 
K magnitude at the long-period end of the sequence for 
fundamental mode pulsators (Miras) \citep{woo98,woo03,lw04}.  
Once the heavy mass-loss phase begins, the period increases 
greatly due to the reduced mass while $M_{bol}$ changes very 
little \citep{vw93}.  Thus the brightening in $M_{bol}$ with 
$\log P$ seen for stars with low mass-loss rates no longer 
holds and the sources in Table 2 no longer fall along a
well-defined sequence, even in a plot of $M_{bol}$ vs. 
$\log P$. 

It should be remembered that there is a selection effect 
operating here which almost certainly acts to limit the 
maximum period seen in the optical databases.  As the stars 
lose mass, their pulsation periods will increase and grow 
more violent, and their mass-loss rate will increase 
\citep[e.g.][]{vw93}.  As a result, their optical magnitudes 
will become fainter due to their thicker dust shells, 
excluding them from the MACHO and OGLE databases.  
Infrared light curves are necessary to determine the periods 
for the reddest stars.

Table 2 also presents the periods of pulsation and the 
bolometric luminosity.  We estimated the bolometric luminosity
by summing the IRS spectrum from 5.1 to 36~\mum\ and also
summing linear interpolations through the photometry from
Siding Spring Observatory.  The photometric points generally
lie at wavelengths affected by molecular absorption bands and
thus lie well below the continuum, requiring a correction
which typically raises the luminosity by $\sim$40\%.
We estimated the strength of this correction from the SWS 
sample of carbon stars (defined in \S 3.2) by summing the
flux from 2.4 to 5.1~\mum, which the IRS does not cover,
and comparing it to the sum from 5.1 to 35~\mum.  The 
resulting correction is a smooth function of [6.4]$-$[9.3] 
color (defined in \S 3.1), and we estimated its strength for 
our SMC targets based on their color.  We also extrapolate a 
Wien distribution to shorter wavelengths and a Rayleigh-Jeans 
tail to the red, but this modification has only a small 
effect on the luminosity.  We have begun to fit our spectra 
using radiative transfer models, and while that work is still 
in its early stages, we can confirm that it produces 
luminosities typically within 10\% of luminosities estimated 
here (the worst discrepancies are 25\%).  For all sources, 
the assumed distance modulus to the SMC is 18.9.

\section{Analysis} 

\subsection{The Manchester method} 

Our analysis is based on the ``Manchester method,'' which is
designed to place the analyses of the samples of carbon stars
presented here and by \cite{mcpw06} on a common basis.  The 
method defines two color indices, [6.4]$-$[9.3] and 
[16.5]$-$[21.5], to characterize the color temperature of the
star-plus-dust system on the blue and red sides of the 
11.3~\mum\ SiC feature.  Amorphous carbon dominates the dust 
emission, and its optical efficiency falls roughly as 
$\lambda^{-1.9}$, making it difficult to distinguish its 
emission from the stellar photosphere.  That task requires 
radiative transfer modelling, which we leave for the future.  
In the following description, ``continuum'' refers to the 
combined spectrum of the star and the amorphous carbon.  This 
method uses a simple scheme of fitting line segments under 
the narrow spectral features to approximate the 
``continuum.'' 

The [6.4]$-$[9.3] color is calculated by separately summing 
the total spectral emission from 6.25 to 6.55 and 9.10 to 
9.50~\mum\ (see Figure 4).  It serves as an indicator of the 
temperature of the carbon-rich dust and its contribution 
relative to the stellar contribution.  In the analysis which 
follows, we use it to discriminate between optically thin, 
warm shells and optically thick, cool shells.  The 6.4~\mum\ 
band measures the continuum between the CO band at 
$\sim$5.0~\mum\ and the C$_2$H$_2$ band at 7.5~\mum, while 
the 9.3~\mum\ band falls between the latter band and an 
absorption band at $\sim$10.0~\mum\ which most likely arises
from C$_3$ \citep{jhl00}. 

The [16.5]$-$[21.5] color, measured over the wavelength 
ranges 16.0--17.0 and 21.0--22.0~\mum, is the basis for 
estimating the cool dust temperature underlying the MgS 
feature, which extends from $\sim$24~\mum\ past the red end 
of the LL spectral range (see Figure 5).  While the IRS data 
provide a substantial improvement in sensitivity over the 
spectra from the SWS on \iso, the wavelength range is more 
limited.  Data from LL are only usable out to $\sim$36~\mum, 
while the SWS provided good coverage out to 45~\mum.  We have
less information about the red side of the MgS feature, but 
we have improved the blue side, where data in Band 3E on the 
SWS (27--28~\mum) proved to be of dubious quality.

Since we cannot observe the continuum to the red of the MgS
feature, we extrapolate the continuum from the blue side 
assuming a Planck function with the temperature corresponding 
to the [16.5]$-$[21.5] color.  The bandpasses for the 16.5 and
21.5~\mum\ measurements were chosen to sample the continuum
between the 13.7~\mum\ C$_2$H$_2$ band and the blue end of the
MgS feature at 24~\mum, and they avoid any potential problems 
at the joint between LL orders 1 and 2 at $\sim$19--21~\mum. 

We measure the strength of the MgS feature by summing the flux 
from 24 to 36~\mum\ above the extrapolated continuum.  Any 
extrapolation method has its dangers, and our measurement of 
the MgS feature is more prone to error than our measurements 
of the C$_2$H$_2$ bands or the SiC feature.  In some of the
spectra, the observed spectrum in the 24--36~\mum\ range 
actually passes {\it under} the extrapolated Planck function,
emphasizing the limits of our method.  In none of these 
cases does there appear to be a MgS dust feature in the 
spectrum.  In other spectra (most notably MSX SMC 066), the 
spectrum continues to climb to the red edge of the LL 
wavelength range, suggesting that the long-wavelength excess 
is more likely a result of background contamination than MgS 
emission.  Only six of the 19 sources in our SMC sample show 
a clear MgS feature with an upturn at 24~\mum\ and a downturn 
at 36~\mum.  In the other cases, we have set the strength of 
the MgS feature to zero.


To measure the strengths of the C$_2$H$_2$ absorption bands 
at 7.5 and 13.7~\mum\ and the SiC emission feature at 
11.3~\mum, we fit line segments to the continuum on either 
side and use this as an estimate of the continuum above or 
below the feature.  For the molecular bands, we report 
equivalent widths (EW), while for the dust features, we 
report the ratio of their integrated flux to the integrated 
continuum flux.  Table 3 provides the wavelengths used to 
estimate the continuum.  The features were integrated between 
the continuum wavelengths.

\begin{figure} 
\includegraphics[width=3.5in]{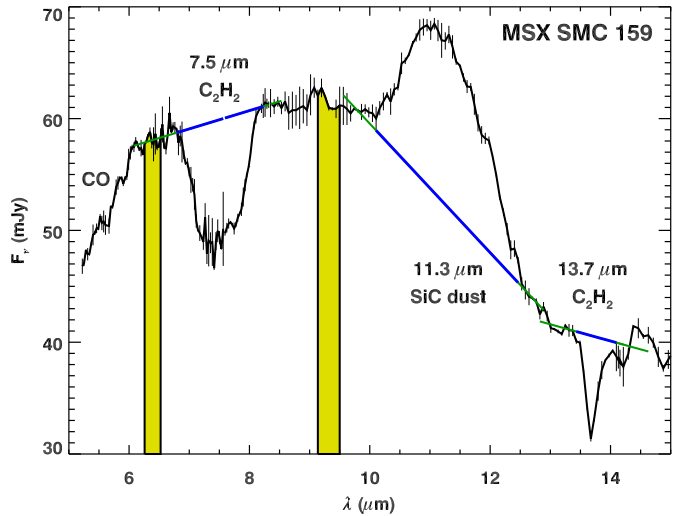}
\caption{An example of the extraction of the molecular bands
and the SiC dust feature from the spectrum of MSX SMC 159
using line segments to approximate a continuum.  The vertical
bars at 6.4 and 9.3~\mum\ show the wavelength ranges used to
determine the [6.4]$-$[9.3] color.}
\end{figure}

\begin{figure} 
\includegraphics[width=3.5in]{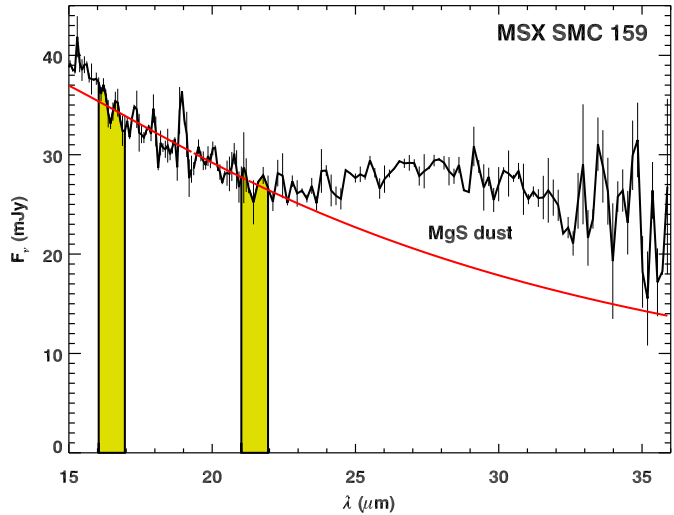}
\caption{The extraction of the strength of the MgS dust
feature, using MSX SMC 159 as an example.  The vertical bars
at 16.5 and 21.5~\mum\ show the wavelength ranges used to
measure the [16.5]$-$[21.5] color, from which we estimate
a color temperature used to extrapolate the continuum
underneath the MgS feature beyond 24~\mum\ (dashed line).}
\end{figure}

\begin{deluxetable}{lccc} 
\tablecolumns{4}
\tablewidth{0pt}
\tablenum{3}
\tablecaption{Fitting wavelengths}
\tablehead{
  & \colhead{$\lambda$} &
  \colhead{Blue continuum} & \colhead{Red continuum} \\
  \colhead{Feature} & \colhead{(\mum)} & \colhead{(\mum)} & \colhead{(\mum)} }
\startdata
C$_2$H$_2$ abs.         &   7.5  &  6.08--6.77  &  8.25--8.55  \\
SiC dust em.            &  11.3  &  9.50--10.10 & 12.50--12.90 \\
C$_2$H$_2$ abs.         &  13.7  & 12.80--13.40 & 14.10--14.70
\enddata
\end{deluxetable}


Figures 4 and 5 illustrate how we determine the colors and
feature strengths using MSX SMC 159 as an example.  For
the SiC feature, we also determine the central wavelength,
which is defined as that wavelength with equal amounts of
flux from the feature to either side after continuum 
subtraction.


\begin{figure}[ht] 
\includegraphics[width=3.5in]{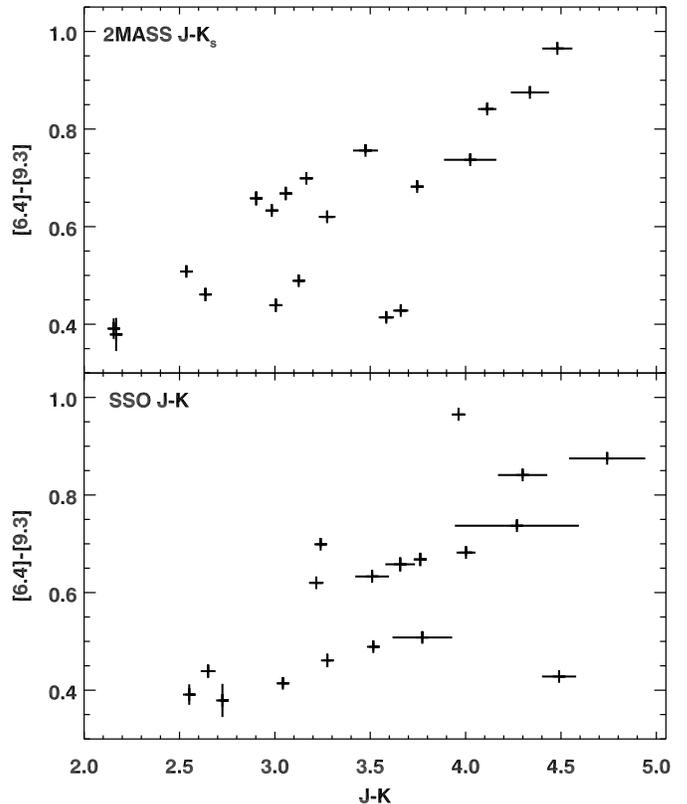}
\caption{The [6.4]$-$[9.3] colors extracted from the spectra,
plotted as a function of J$-$K color.  The top panel plots the 
2MASS J$-$K$_s$ color, and the bottom panel plots the SSO
J$-$K color.  In both cases, the [6.4]$-$[9.3] color correlates
reasonably well with the near-infrared color.  The overall 
shift between the two J$-$K colors results from the shift in 
the position of the J and K bands between the two systems.  The 
two most notable outliers at J$-$K$_s$$\sim$3.6 and 
[6.4]$-$[9.3]$\sim$0.4 are MSX SMC 066 and 200.  MSX SMC 200 is 
the only outlier when looking at SSO J$-$K color.}
\end{figure}

Figure 6 compares the [6.4]$-$[9.3] color measured from
the spectra with the J$-$K color using both the 2MASS and SSO
photometry.  The apparent overall shift in J$-$K between the 
2MASS and SSO systems is real; it arises from a shifts in the
band centers.  In the 2MASS system, the centers of the J and 
K$_s$ bands are 1.25 and 2.15~\mum, respectively.  In the
SSO system, the J and K bands are shifted to 1.24 and 2.22~\mum.
The carbon stars in this sample are very red and far from 
Rayleigh-Jeans distributions, which accentuates the effect of 
moving the band centers.

The [6.4]$-$[9.3] color shows a correlation with J$-$K in both 
systems.  The J$-$K color is more sensitive to photospheric 
emission, while the [6.4]$-$[9.3] colors is more sensitive to 
emission from the dust, so the scatter in the data is not 
surprising, even though the SSO photometry was obtained at the 
same phase as the spectra used to determine the [6.4]$-$[9.3] 
color.  The two most significant outliers are 
MSX SMC 066 and 200.  Both lie at J$-$K$_s$$\sim$3.6 and 
[6.4]$-$[9.3]$\sim$0.4.  In the comparison with the SSO J$-$K 
color, only MSX SMC 200 remains an outlier.  


Table 4 presents our measurements of the sample of 19 carbon
stars in the SMC.  We do not report a MgS strength unless we
believe it be valid, and we do not report a central wavelength
for SiC features too weak to be measured.  Figures 7 and 8 plot
the strengths of the molecular absorption bands and dust
emission features, respectively, as a function of the
[6.4]$-$[9.3] color.

\begin{deluxetable*}{lcccccccc} 
\tablenum{4}
\tablecolumns{9}
\tablewidth{0pt}
\tabletypesize{\footnotesize} 
\tablecaption{Spectroscopic Results}
\tablehead{
  \colhead{MSX} & \colhead{[6.4]$-$[9.3]} & \colhead{[16.5]$-$[21.5]} &
  \colhead{EW at 7.5~\mum}         & \colhead{EW at 13.7~\mum} &
  \colhead{$\lambda_{\rm SiC}$}   & \colhead{} &
  \colhead{ }         & \colhead{Infrared} \\
  \colhead{SMC}    & \colhead{(mag)}  & \colhead{(mag)} &
  \colhead{(\mum)} & \colhead{(\mum)} & \colhead{(\mum)} &
  \colhead{SiC/continuum}      & \colhead{MgS/continuum}      & \colhead{Spec. Class}
}
\startdata
033 &  0.49$\pm$0.01 & -0.06$\pm$0.03  & 0.053$\pm$0.004 & 0.054$\pm$0.003 &
      11.34$\pm$0.06 & 0.036$\pm$0.003 & \nodata         & 2.CE \\
036 &  0.76$\pm$0.01 &  0.17$\pm$0.02  & 0.244$\pm$0.005 & 0.066$\pm$0.007 &
      11.26$\pm$0.03 & 0.088$\pm$0.004 & 0.211$\pm$0.047 & 3.CE \\
044 &  0.51$\pm$0.01 & -0.09$\pm$0.03  & 0.043$\pm$0.006 & 0.036$\pm$0.004 &
      11.25$\pm$0.09 & 0.022$\pm$0.005 & \nodata         & 2.CE \\
054 &  0.74$\pm$0.01 &  0.16$\pm$0.02  & 0.187$\pm$0.004 & 0.057$\pm$0.004 &
      11.27$\pm$0.02 & 0.179$\pm$0.004 & 0.116$\pm$0.027 & 3.CE \\
060 &  0.96$\pm$0.00 &  0.31$\pm$0.02  & 0.116$\pm$0.004 & 0.030$\pm$0.004 &
      11.42$\pm$0.04 & 0.051$\pm$0.003 & 0.253$\pm$0.032 & 3.CR \\
062 &  0.67$\pm$0.01 &  0.24$\pm$0.02  & 0.108$\pm$0.007 & 0.067$\pm$0.009 &
      11.23$\pm$0.07 & 0.066$\pm$0.005 & \nodata         & 3.CE \\
066 &  0.41$\pm$0.01 &  0.17$\pm$0.03  & 0.084$\pm$0.002 & 0.058$\pm$0.005 &
      11.34$\pm$0.04 & 0.034$\pm$0.002 & \nodata         & 2.CE \\
091 &  0.66$\pm$0.01 &  0.15$\pm$0.04  & 0.241$\pm$0.012 & 0.107$\pm$0.013 &
      11.30$\pm$0.05 & 0.148$\pm$0.010 & \nodata         & 3.CE \\
093 &  0.44$\pm$0.01 &  0.23$\pm$0.05  & 0.257$\pm$0.006 & 0.032$\pm$0.006 &
      11.06$\pm$0.16 & 0.025$\pm$0.006 & \nodata         & 2.CE \\
105 &  0.84$\pm$0.01 &  0.25$\pm$0.02  & 0.186$\pm$0.007 & 0.044$\pm$0.005 &
      11.22$\pm$0.02 & 0.154$\pm$0.006 & 0.253$\pm$0.028 & 3.CR \\
142 &  0.38$\pm$0.03 & -0.23$\pm$0.22  & 0.200$\pm$0.033 & 0.058$\pm$0.038 &
      \nodata        & 0.018$\pm$0.018 & \nodata         & 2.CE \\
159 &  0.87$\pm$0.01 &  0.31$\pm$0.02  & 0.163$\pm$0.003 & 0.061$\pm$0.005 &
      11.21$\pm$0.02 & 0.178$\pm$0.004 & 0.427$\pm$0.027 & 3.CR \\
162 &  0.46$\pm$0.01 &  0.06$\pm$0.12  & 0.131$\pm$0.010 & 0.015$\pm$0.010 &
      11.31$\pm$0.09 & 0.059$\pm$0.007 & \nodata         & 2.CE \\
163 &  0.68$\pm$0.01 &  0.17$\pm$0.01  & 0.140$\pm$0.005 & 0.056$\pm$0.004 &
      11.27$\pm$0.02 & 0.163$\pm$0.004 & 0.148$\pm$0.019 & 3.CE \\
198 &  0.62$\pm$0.01 &  0.09$\pm$0.02  & 0.115$\pm$0.007 & 0.071$\pm$0.005 &
      11.19$\pm$0.05 & 0.049$\pm$0.003 & \nodata         & 3.CE \\
200 &  0.43$\pm$0.01 & -0.11$\pm$0.03  & 0.040$\pm$0.004 & 0.066$\pm$0.005 &
      11.44$\pm$0.18 & 0.009$\pm$0.003 & \nodata         & 2.CE \\
202 &  0.39$\pm$0.02 &  0.06$\pm$0.04  & 0.134$\pm$0.007 & 0.041$\pm$0.009 &
      \nodata        & 0.005$\pm$0.008 & \nodata         & 2.CE \\
209 &  0.70$\pm$0.01 &  0.12$\pm$0.03  & 0.097$\pm$0.005 & 0.050$\pm$0.003 &
      11.16$\pm$0.05 & 0.028$\pm$0.002 & \nodata         & 3.CE \\
232 &  0.63$\pm$0.01 &  0.07$\pm$0.03  & 0.088$\pm$0.005 & 0.069$\pm$0.005 &
      10.98$\pm$0.09 & 0.028$\pm$0.003 & \nodata         & 3.CE
\enddata
\end{deluxetable*}

\begin{figure} 
\includegraphics[width=3.5in]{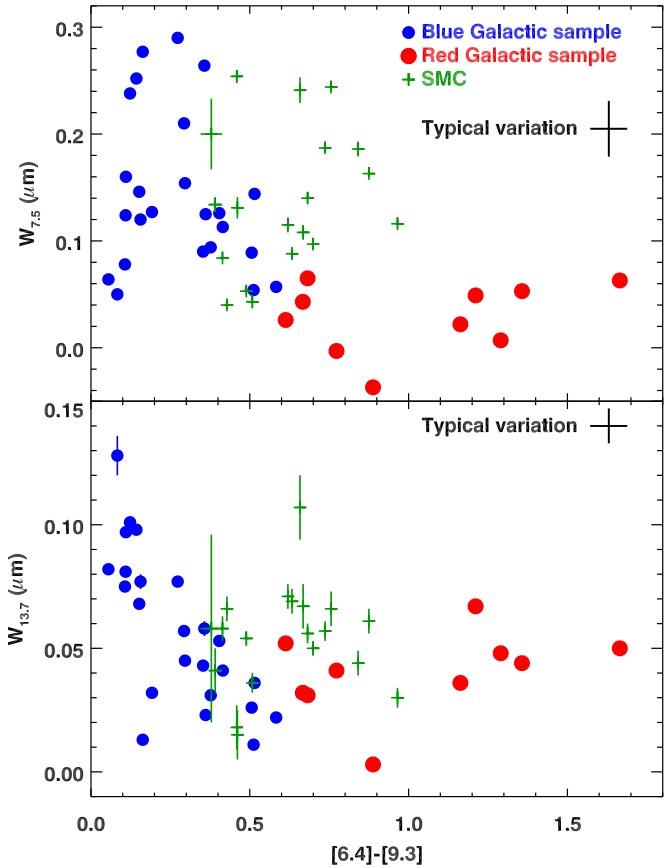}
\caption{The equivalent widths of the C$_2$H$_2$ bands
at 7.5 and 13.7~\mum\ as a function of [6.4]$-$[9.3] color
for both the SMC sample (plus signs) and the SWS Galactic
sample (circles).  The redder sources in the SMC sample
show stronger absorption bands than their counterparts
in the Galaxy.  The black cross labelled ``typical variation''
reflects the expected impact on the data from the variability
of the star, based on an analysis of those sources observed
multiple times by the SWS.}
\end{figure}

\begin{figure} 
\includegraphics[width=3.5in]{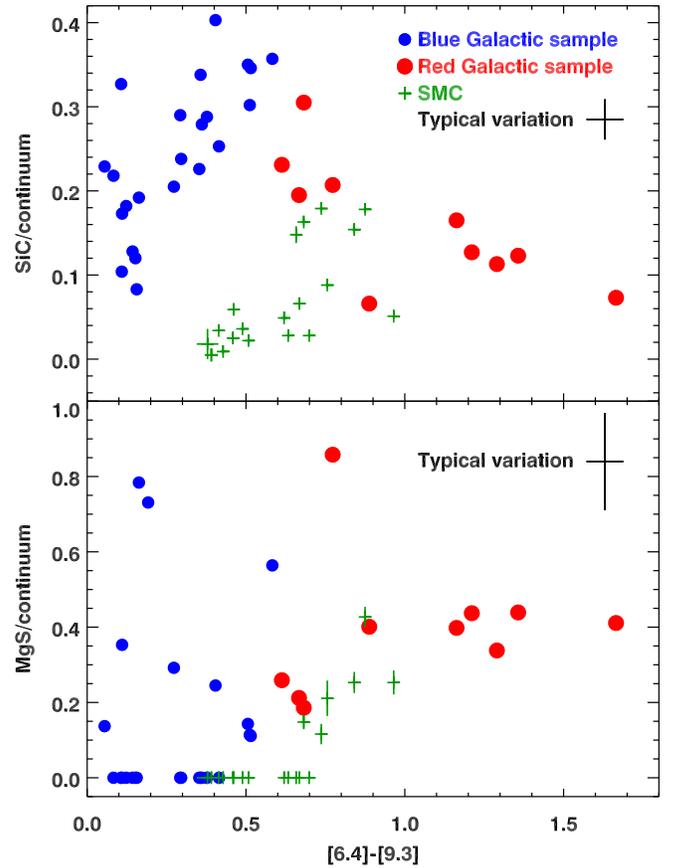}
\caption{The strength of the dust emission features from
SiC at 11.3~\mum\ and MgS at $\sim$30~\mum\ as a function 
of [6.4]$-$[9.3] color.  Symbols are as in Fig. 7.  The SMC 
sample tends to show weaker dust emission than the Galactic 
sample.}
\end{figure}

\subsection{Control sample} 

We have also analyzed the sample of carbon stars in the Galaxy
observed by the SWS using the same method as described above.
This sample includes a total of 63 spectra of 39 sources,
consisting of 29 sources classified by \cite{kra02} as 2.CE 
(warm carbon-rich dust emission), four classified as 3.CE 
(cool carbon-rich dust emission), and six classified as 3.CR 
(cool and reddened carbon-rich dust emission).  Two sources 
(S Cep and V460 Cyg) were observed twice, and four sources 
were observed several times each (V CrB, 6 times; V Cyg, 6; 
T Dra, 8; R Scl, 6).  Five of the 2.CE spectra were dropped 
from the sample because they are too noisy for meaningful 
analysis.

In this paper, we concentrate on the properties of the SWS 
sample as a control group for our SMC sample.  We will present 
a more detailed analysis of the Galactic sample elsewhere 
(Sloan et al. in preparation).  \cite{ych04} have examined 
most of the spectra in this sample, and they have developed a 
classification system similar to ours, running from A to D 
for increasingly red sources.  Hereafter, we will refer to 
the 24 2.CE sources as the blue Galactic sample, and the 10 
3.CE and 3.CR sources as the red Galactic sample. 

The boundary between the 2.CE and 3.CE sources in the 
Galactic sample lies at [6.4]$-$[9.3]=0.60, and similarly, a 
color of 0.80 separates the 3.CE and 3.CR sources.  While the 
earlier classification of the SWS data \citep{kra02} did not 
make use of the [6.4]$-$[9.3] color, we apply it here to 
distinguish these classes in the SMC sample.  Table 4 
includes these classifications.

For the sources observed multiple times, we generated and
analyzed a co-added spectrum.  We also examined the 
individual spectra to investigate how much the spectral
properties of a given source can vary over a pulsation
cycle.  For each source, we determined the standard
deviation of the various measured properties and from these
determined a median variation, which appears in the relevant 
figures (7--10) as a black cross labelled ``typical 
variation.''

\section{Results} 

\subsection{Molecular band strengths} 


Figure 7 plots the equivalent widths of the C$_2$H$_2$ bands
as a function of [6.4]$-$[9.3] color for both the SMC sample
and the Galactic sample.  At first thought, it might seem
possible that the lower metallicities in the SMC would lead 
to optically thinner dust shells, which would make it easier 
to see through the dust envelope into the molecular absorption
zone.  We have accounted for this effect by plotting the band 
strengths as a function of [6.4]$-$[9.3] color, which serves 
as a proxy for the optical depth of the shell.  For both the
7.5 and 13.7~\mum\ bands, the two groups overlap significantly
for [6.4]$-$[9.3] $<$ 0.6, but for thicker dust shells, the 
bands are stronger in the SMC sources.

In even the reddest sources in our Galactic sample, where the
line of sight to the stellar photosphere should be veiled by 
the overlying dust, the molecular bands are present.  It is 
unlikely that we are seeing down to the stellar photosphere in 
these cases; it is more likely that the absorbing molecules 
are in an extended envelope well above the stellar photosphere 
and overlapping with the circumstellar dust shell.  Van Loon et 
al. (2006) use a similar argument to explain the strength and
shape of the 3.1 and 3.8~\mum\ bands due to C$_2$H$_2$ 
in a sample of carbon stars in the LMC.

The difference in band strength between the SMC and Galactic
samples is more readily apparent at 7.5~\mum\ than at 
13.7~\mum, probably because we are only measuring the narrow
Q branch of the 13.7~\mum\ feature.  For high column densities, 
the Q branch will be saturated, and one would need to measure 
the broad P and R branches to determine the column density, but
these branches are difficult to separate from the continuum.
It should also be kept in mind that the 7.5 and 13.7~\mum\
bands have different excitation energies, meaning that they
sample different regions of the extended molecular envelope.

\subsection{The SiC dust emission feature} 


The SiC dust emission feature at 11.3~\mum\ lies between two
molecular absorption bands, complicating our ability to
accurately extract its strength.  To the blue, we have the
10.0~\mum\ absorption band attributed to C$_3$ \citep{jhl00}.
This feature does not appear to be strong enough to have a
significant impact on our spectra.  The red side of the SiC 
feature is more of a problem, since the dust emission overlaps 
the P branch of the $\nu_5$ band, which in some cases could 
extend past $\sim$12.5~\mum\ \citep{ato99}.  If we fit the 
continuum too far to the red, then we are really fitting
molecular absorption.  Fitting too far to the blue cuts off
dust emission.  We have chosen the 12.5--12.9~\mum\ range as 
a compromise.  In our analysis, we have experimented with 
other wavelength ranges, and we have found that the 
conclusions we present below are robust in a qualitative 
sense.  Some of the quantitative results, of course, do 
depend on where we fit the continuum.

The top panel of Figure 8 compares the strength of the SiC
dust emission features in the Galactic sample and in the SMC. 
Five sources in the SMC, all with [6.4]$-$[9.3] $>$ 0.6, show
SiC features stronger than 10\% of the integrated continuum,
similar in strength to their Galactic counterparts with the
same colors.  The remaining sources in the SMC sample, 
especially the blue sources, show comparatively weaker SiC
features.

The apparent trend of decreasing SiC emission strength as a 
function of [6.4]$-$[9.3] color in the red Galactic sample 
must be considered with caution.  While amorphous carbon 
dominates the dust in these shells, enough SiC dust can be 
present to force the 11.3~\mum\ feature into absorption,
as first suggested by \cite{jon78} and supported by the
comparison of laboratory data with both ground-based 
observations \citep{spe97} and {\it ISO}/SWS data 
\citep{cle03}.  Thus, the apparent weakening of the SiC 
feature in the Galactic sample as the spectra get redder 
could in fact result from self-absorption in the feature.

The SiC feature is notorious for appearing at a wide range
of wavelengths.  While the emission feature typically appears
at 11.3~\mum, it can be seen out to 11.7~\mum.  In extreme
carbon stars, the feature can appear in absorption, usually
at $\sim$10.8~\mum.  \cite{spe05} have presented recent 
laboratory measurements to explain these wavelength shifts.  
SiC grains emit with two components, a longitudinal optic 
(LO) at 10.8~\mum\ and a transverse optic (TO) to the red.  
As the SiC grains increase in size, the LO increases with 
respect to the TO, shifting the emission feature to the blue.  
\cite{spe05} explain the absorption at 10.8~\mum\ as the 
result of different grain size distributions, with the cool 
absorbing component containing larger grains with an enhanced 
LO and the warmer emitting component containing smaller 
grains with an enhanced TO, shifting the feature from 
11.3~\mum\ to the red.  If this scenario is valid, we 
can use wavelength to track the optical depth of the SiC 
feature.


\begin{figure}
\includegraphics[width=3.5in]{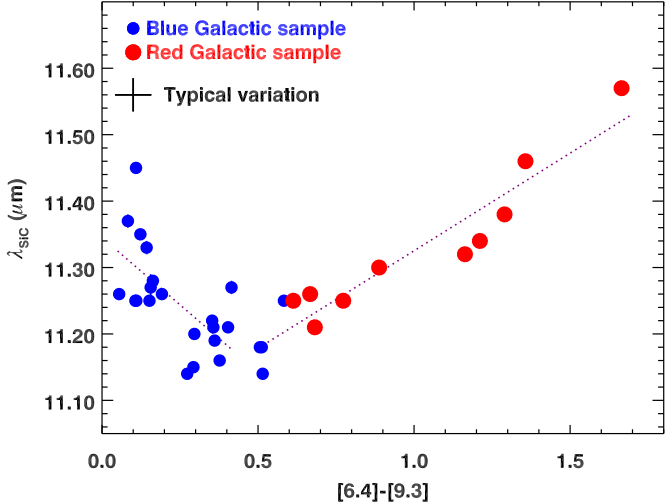}
\caption{The apparent central wavelength of the SiC feature 
as a function of [6.4]$-$[9.3] color for the Galactic sample. 
Symbols are as defined in Fig. 7.  The wavelength decreases
from colors of 0.0 to 0.5, then increases steadily to almost
11.6~\mum\ for the reddest source in our sample.  The 
increase in apparent central wavelength for the red sources 
is consistent with the stronger self-absorption at 10.8~\mum\ 
described by \cite{spe05}.  Line segments fitted to the data 
on either 
side of [6.4]$-$[9.3]=0.45 are shown as dotted lines.}
\end{figure}

Figure 9 illustrates how the apparent central wavelength of 
the SiC feature varies as a function of [6.4]$-$[9.3] color 
for the Galactic SWS sample.  For colors from 0.0 to 0.5, the 
apparent central wavelength shows a great deal of scatter 
between values of 11.1 and 11.5~\mum, but a general trend of 
decreasing wavelength with reddening color is noticeable.  
The Pearson correlation coefficient \citep[e.g.][]{pr88} of 
a line segment fit to the blue of [6.4]$-$[9.3]=0.45 is 
0.63, indicating a reasonable correlation despite the scatter.  
Beyond [6.4]$-$[9.3]=0.5, the apparent central wavelength 
increases monotonically to nearly 11.6~\mum\ for the reddest 
source in our SWS sample.  The correlation coefficient in 
this range is an impressive 0.95.  

In Figure 8 (top), the measured strength of the SiC emission 
generally increases with [6.4]$-$[9.3] for the blue Galactic 
sources (although there are a few notable outliers).  The 
apparent turnover at [6.4]$-$[9.3]$\sim$0.5 and the 
subsequent drop in apparent strength in the Galactic sample
can now be interpreted not as an intrinsic decline in SiC 
dust abundance, but as the growing effect of self-absorption 
on the measured strength.



\begin{figure} 
\includegraphics[width=3.5in]{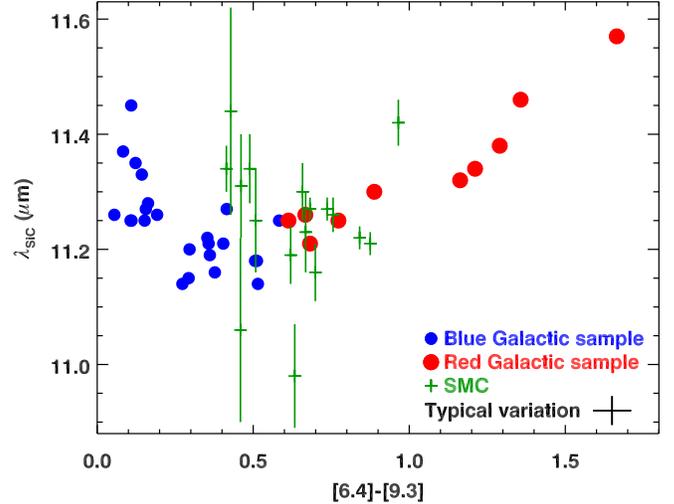}
\caption{A comparison of the apparent central wavelength of 
the SiC feature for the Galactic and SMC samples, as in Fig. 
9.  Symbols are as in Fig. 7.  The SMC sample occupies a 
tight wavelength range between 11.15 and 11.45~\mum, with 
the exception of MSX SMC 093 and 232, which both show SiC
features shifted to the blue.  Few of the SMC sources show 
evidence of self-absorption in the SiC feature, which is 
consistent with the intrinsic weakness of this feature in 
the SMC sample compared to the Galactic sample.}
\end{figure}

Figure 10 compares the apparent central wavelength of the SiC 
feature ($\lambda_{SiC}$) in the Galactic and SMC samples.  Two
sources, MSX SMC 060 and 200, show central wavelengths $>$
11.4~\mum, but in MSX SMC 200, the uncertainty is nearly 0.2~\mum.
MSX SMC 060 has a central wavelength of 11.42~\mum\ and a 
[6.4]$-$[9.3] color of 0.97.  The strength of the SiC feature in 
its spectrum is lower than some of the bluer sources, making it 
probable that this source shows some mild effects from 
self-absorption.  The remaining sources in the sample show no 
evidence for self absorption, leading us to conclude that as a
group, the SMC spectra do indeed show less emission from SiC 
dust than their Galactic counterparts.

Even where the two samples show similar apparent strengths of
the SiC feature, the Galactic sources have features shifted
more to the red on average, indicating that they are more
self-absorbed.  Correcting for the effect of self-absorption
on our measurements would make the two samples even more
distinct.


\begin{figure} 
\includegraphics[width=3.5in]{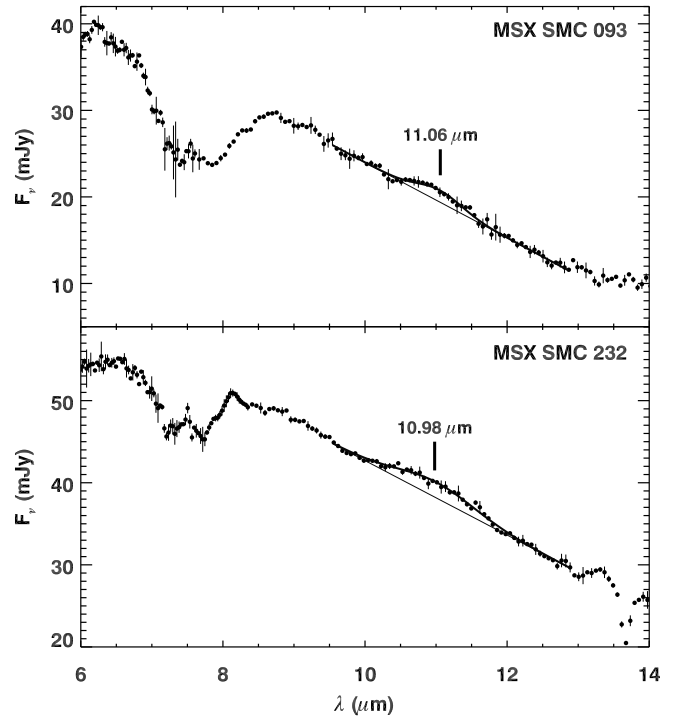}
\caption{The spectra of MSX SMC 093 and 232, showing the weak
and blue-shifted SiC features and the Gaussians fit to them.  
The data are plotted as filled circles with error bars, the 
fitted linear continua under the 11.0~\mum\ emission features 
appear as thin lines, and the Gaussian fits as thick lines.  
The line centers, as marked, are at 11.06 and 10.98 for MSX 
SMC 093 and 232, respectively, while the Gaussians are 
centered at 11.05 and 11.00~\mum, respectively.}
\end{figure}

Two of the sources, MSX SMC 093 and 232, are outliers in Figure
10, showing blue-shifted emission features peaking at 11.06 
and 10.98~\mum, respectively.  Figure 11 shows these spectra, 
along with Gaussians fit to the features.  The Gaussians are 
centered at 11.05~\mum\ for MSX SMC 093 and 11.00~\mum\ for 
MSX SMC 232.  Because these features are weak compared to
the continuum, their apparent position is somewhat sensitive
to how we choose to fit the continuum.  As an experiment, we
changed the wavelength range for continuum fitting on the red
side of the SiC feature to 12.0--12.9~\mum.  While the 
positions in the spectra of MSX SMC 093 and 232 remained within
0.02~\mum, the SiC feature in MSX SMC 209 moved to 
11.09$\pm$0.05~\mum\ (from 11.16~\mum), which would make it a
third outlier in the sample.

Both MSX SMC 093 and 232 are among the four sources with the 
weakest measured SiC features.  (MSX SMC 209 is as well.)  
Following the lead of \cite{spe05}, we hypothesize that the 
emission from these grains is dominated by the longitudinal 
optic (LO), which indicates that the grains are larger than 
usually seen.  Thus, these spectra show the enigmatic 
combination of both fewer and larger SiC grains. 

A similar process might explain the overall trend apparent in
the blue Galactic sources of decreasing central wavelength
until the [6.4]$-$[9.3] color reaches $\sim$0.5.  In this 
case, the trend of increasing grain size with increasingly 
red colors and thus growing optical depth makes more sense.

\subsection{The MgS dust emission feature} 

Only one third of our sample shows an unambiguous MgS 
feature.  Some spectra (such as MSX SMC 066, 142, and 209), 
appear to be contaminated by background emission which would 
hide any MgS feature, but the majority of the remaining 12 
spectra simply show no evidence of MgS emission above the
continuum extrapolated from the 16--22~\mum\ region.

None of the SMC spectra with [6.4]$-$[9.3]$<$0.75 show a MgS
feature, while at least some Galactic spectra show a feature
throughout the range of colors.  For the six spectra which 
do show MgS features, five have weaker features than their 
Galactic counterparts with similar colors.  While the SMC 
spectra are generally noisy in this spectral range and 
sometimes contaminated by background emission, and our
extrapolation method does have its limits, it still seems 
that the MgS feature is weaker in the SMC sample than in the 
Galactic sample.


\section{Discussion} 

\subsection{Dust emission and abundances} 

Both the SiC and MgS dust emission features tend to be 
weaker in the carbon stars in the SMC than in the Galaxy.
The most straightforward explanation for these differences 
is that they result from the lower metallicity in the SMC.
Mg, Si, and S are all products of $\alpha$-capture, and
studies of halo stars show that the abundances of these 
elements behave similarly, decreasing as iron abundance 
decreases \citep[See the review by][]{wst89}.  

The sample of carbon stars in the SMC divides easily into 
three groups.  Eight sources, which we call the blue sources, 
have 0.3$<$[6.4]$-$[9.3]$<$0.6, and all show weak SiC 
emission.  The red sources (0.6$<$[6.4]$-$[9.3]$<$1.0), 
separate into two groups, six sources with weak SiC features, 
and five with strong SiC features.  Table 5 compares the mean
strength of the SiC features and mean color in these groups
to groups from the Galactic sample with the same color 
ranges.  The Galactic sample does not show a dichotomy in 
SiC strength among the red sources like the SMC sample.


\begin{deluxetable}{lrcc} 
\tablecolumns{4}
\tablewidth{0pt}
\tablenum{5}
\tablecaption{Mean SiC strengths}
\tablehead{ \colhead{Sample} & \colhead{N} & 
  \colhead{$<$[6.4]$-$[9.3]$>$} & \colhead{$<$SiC/continuum$>$} }
\startdata
Galactic blue        & 10 & 0.44 & 0.314$\pm$0.054 \\
SMC blue             &  8 & 0.44 & 0.026$\pm$0.017 \\ \\
Galactic red         &  5 & 0.73 & 0.201$\pm$0.087 \\
SMC red, weak SiC    &  6 & 0.72 & 0.052$\pm$0.023 \\
SMC red, strong SiC  &  5 & 0.76 & 0.164$\pm$0.014 
\enddata
\end{deluxetable}

The strength of the SiC feature in the SMC sample, compared
to the Galactic sample, is down by a factor of 12 for the blue
sources, 4 for the red sources with weak SiC features, and
1.2 for the red sources with a strong SiC feature.  Our
measurements of the SiC feature in the red sources in the
Galactic sample may be underestimated due to self-absorption,
so the difference between the two samples may actually be larger.
Abundances of Mg and Si in the SMC are about 15\% of their 
Solar values, based on measurements of the H II region N88a 
\citep{ku99} and B supergiants \citep{le03,tr04,du05}.  Our 
measured dust strengths are roughly consistent with these
abundance differences.  The lower SiC strengths in the bluer
sources could be explained if they representd an older and
more metal-poor sample.

\subsection{The strong SiC sources} 

The five sources in the SMC sample which show SiC emission 
nearly as strong as their Galactic counterparts are MSX SMC 
054, 091, 105, 159, and 163.  Four of these sources are also 
among the six showing a measurable MgS feature; MSX SMC 091 
is the exception.  Conversely, MSX SMC 036 and 060 are the 
two MgS sources which don't show strong SiC bands.
As Figure 8 shows, the five strong SiC sources are all
in the redder half of the sample in [6.4]$-$[9.3] color.
The MgS sources are grouped even more clearly; the six
reddest sources all show MgS features.  


Figure 12 (top) shows that the strong SiC sources also tend 
to have the reddest J$-$K colors.  Four of these sources are 
among the five reddest.  MSX SMC 091, again, is the 
exception, although it is still redder than half the sample.  
The one red source with J$-$K $>$ 4 but without strong SiC 
emission is MSX SMC 200.  Figure 12 uses the photometry from 
SSO.  The 2MASS photometry from an earlier epoch gives 
similar results, except that MSX SMC 060 is the one weak 
SiC source among the reddest five in J$-$K.  We have noted 
above that MSX SMC 060 and 200 have the most red-shifted SiC 
features in our sample; we could possibly be underestimating 
the strength of these features due to self-absorption.  The 
middle panel of Figure 12 shows that the strong SiC sources 
are also among the redder sources in K$-$L, although the 
trend is not as obvious as it is in J$-$K.
  
Given that the weakness of the SiC and MgS features in the 
general sample is consistent with the expected lower 
abundances of $\alpha$-capture products in the SMC, it is 
conceivable that the sources with stronger SiC features 
differ from the rest of the SMC sample by having enhanced 
metallicities.  If the metallicities resulted from
comparative youth, then they might be located near younger
clusters in the SMC, but there is nothing significant about
their locations in the SMC.  Younger stars would have a 
higher mass, and on average, they might show a higher 
luminosity.  However, as the bottom panel of Figure 12 
shows, only one of the eight sources with a bolometric 
luminosity $>$ 10$^4$ L$_{\sun}$ has a strong SiC feature.
The strong SiC sources cover a wide range of luminosities, 
although four of the five tend to cluster in the lower 
half of the luminosity range.

The only clear distinctions between the strong SiC sources
and the remaining carbon stars are that they (1) tend to
also show MgS dust and (2) tend to be redder than the 
remaining sample at J$-$K and to a lesser degree, K$-$L and
[6.4]$-$[9.3].  These redder colors imply a greater amount
of amorphous carbon in their dust shells.

\begin{figure} 
\includegraphics[width=3.5in]{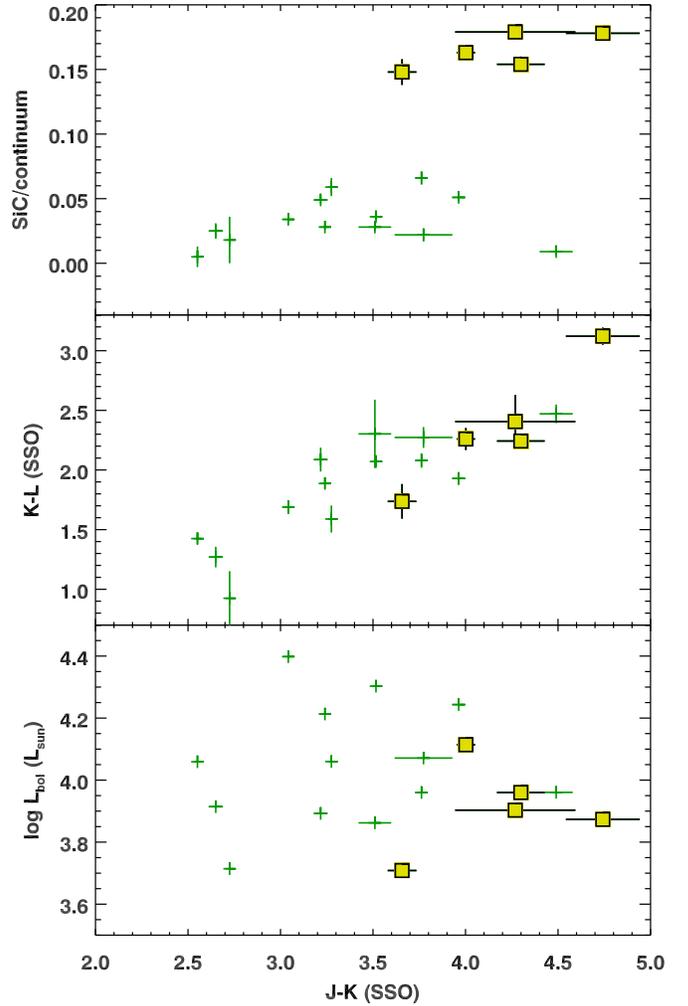}
\caption{The strength of the SiC dust emission as a 
function of J$-$K color ({\it top}), our sample plotted 
on a JKL color-color diagram ({\it middle}), and 
bolometric luminosity vs. J$-$K color ({\it bottom}).  
The strong SiC sources are plotted with boxes.  The
photometry is from Siding Spring Observatory (SSO), 
although it should be noted that the dependence on 2MASS 
J$-$K$_s$ colors is similar.  The strong SiC sources tend to 
be among the redder sources in our sample, especially at 
J$-$K.  \S 2.4 explains how we estimated the bolometric 
luminosity.  The strong SiC sources are not among the 
brightest sources in the sample.}
\end{figure}

\subsection{Other unusual sources} 

The star MSX SMC 060 shows a rather unusual lightcurve 
(Figure 3), and it has the reddest [6.4]$-$[9.3] color in our
sample of carbon stars.  It shows evidence for a long period 
of about 450--600 days, and when it is bright it shows a 
typical semi-regular (low overtone) pulsation with a period 
of 70--85 days.  In some ways, this star is like an R CrB 
star, undergoing episodic dimming events, a slow recovery, 
and then showing semi-regular pulsation when it is optically 
bright.  Photometrically, it is certainly as red in J$-$K and 
[6.4]$-$[9.3] as the two R CrB stars in the SMC observed with 
the IRS \citep{kra05}, MSX SMC 014 and 155.  They have 
[6.4]$-$[9.3] = 0.92 and 0.77, respectively, compared to 0.96 
for MSX SMC 060.  The similarities in the light curve and 
photometry are interesting, even though the infrared spectrum 
of MSX SMC 060 differs significantly from the R CrB stars, 
which show relatively featureless spectra, while MSX SMC 060 
has clear acetylene bands, strong MgS dust emission, and 
moderate (or possibly self-absorbed) SiC dust emission.  

Two sources in Figure 6 do not follow the general relation
between [6.4]$-$[9.3] and J$-$K$_s$ colors, showing bluer 
colors at [6.4]$-$[9.3] than expected from the J$-$K colors 
in the 2MASS epoch.  They are MSX SMC 066 and 200, and they 
have J$-$K$_s$ colors in the 2MASS epoch of 3.58 and 3.66, 
respectively.  In the SSO photometry, MSX SMC 066 has dropped 
to J$-$K = 3.04, among the bluest colors in our sample, while 
MSX SMC 200 has reddened to 4.49, among the reddest colors in 
our sample.  The variability of MSX SMC 066 might explain why 
it stands out in Figure 6, but MSX SMC 200 has remained red 
at J$-$K, making its blue [6.4]$-$[9.3] color more difficult 
to explain.

The two sources illustrated in Figure 11, MSX SMC 093 and 
232, are outliers in Figure 10 because their SiC features 
have shifted from $\sim$11.3~\mum\ to $\sim$11.0~\mum.  As 
explained in \S 4.2, this shift may arise from an increased 
size in the emitting grains, even though the SiC dust is 
more optically thin.

\subsection{Acetylene and amorphous carbon} 

\subsubsection{C/O ratio} 

The C$_2$H$_2$ absorption bands are stronger in the SMC 
sample than in the Galactic sample, indicating that these 
molecules are more abundant in the atmospheres and outflows 
of carbon stars in the SMC.  This result reinforces previous 
results based on 3--4-\mum\ spectra of carbon stars in the 
Magellanic Clouds and the Sagittarius Dwarf Spheroidal galaxy 
\citep{vl99,mat02,mat05}.  These papers suggest that the 
increased abundance of C$_2$H$_2$ results from a higher C/O 
ratio in low-metallicity carbon stars, which arises for two 
reasons.  First, as the metallicity drops, the dredge-ups 
will grow more efficient \citep{woo81}, injecting more carbon 
into the atmosphere.  Second, low-metallicity AGB stars have 
a lower oxygen abundance to begin with, increasing the C/O 
ratio for a given amount of dredged up carbon \citep{lw04}.


The analysis in this paper uses the [6.4]$-$[9.3] color as 
a proxy for the relative contributions of emission from the 
amorphous carbon dust and the stellar photosphere.
Figure 7 shows that at a given [6.4]$-$[9.3] color, carbon 
stars in the SMC have more acetylene absorption than their 
Galactic counterparts.  Thus, if our assumptions are 
correct, the carbon stars in the SMC have more acetylene
around them than carbon stars {\it with the same amount of 
dust} in the Galaxy. 

\subsubsection{Dust formation efficiency} 

\begin{deluxetable}{lrcc} 
\tablecolumns{4}
\tablewidth{0pt}
\tablenum{6}
\tablecaption{Mean acetylene strengths in the SMC}
\tablehead{ \colhead{ } & \colhead{ } &
  \colhead{$<$EW at 7.5~\mum$>$} & \colhead{$<$EW at 13.7~\mum$>$} \\
  \colhead{Sample} & \colhead{N} & \colhead{(\mum)} & \colhead{(\mum)} }
\startdata
weak SiC    &  6 & 0.141$\pm$0.069 & 0.053$\pm$0.017\\
strong SiC  &  5 & 0.183$\pm$0.038 & 0.064$\pm$0.024
\enddata
\end{deluxetable}

\cite{atb89} suggested acetylene as a probable building
block of aromatic hydrocarbons in circumstellar shells, and
numerical models support this mechanism \citep{ff89,ch92}.
Carbonaceous grains with a mixture of aliphatic and
aromatic bonds can also be produced from acetylene, as shown
by \citet[][ 2005]{kov03}.  The acetylene probably does not
condense directly into solid material.  A nucleation site
provides a surface for the formation of carbonaceous solids 
in the outflows from carbon stars, and \cite{fr89} and 
\cite{cad94} have identified SiC as the most likely 
nucleation site.

Our observations of the SMC reveal less SiC and MgS in the 
dust compared to the Galactic sample, raising another possible
explanation for the excess acetylene seen in the SMC spectra
besides the higher expected C/O ratios.  The lower metallicity
in the SMC leads to lower abundances of heavier elements such 
as Si, Mg, and S and fewer related dust grains.  With less 
SiC to form the nuclei for the condensation of carbon, the dust 
formation may be less efficient, which might in turn explain 
the deeper acetylene absorption for a given amount of 
carbon-rich dust in the SMC (Figure 7).

While this scenario seems plausible, the five strong SiC
sources in our sample do not follow the expected trend.
Since the SiC emission in these sources is nearly as strong 
as the Galactic sources, one might expect more efficient dust 
formation to leave less acetylene compared to the sources 
with weaker SiC features.  Table 6 compares the mean band
strengths for the acetylene in the strong SiC sources to the
same control sample of weak SiC sources in Table 5.
The distributions of the molecular band 
strengths overlap significantly.  If anything, the strong SiC 
emitters have stronger absorption bands.  While the five 
strong SiC sources do not seem to support the scenario we 
propose, one must keep in mind that we have been unable to 
determine why these sources have so much SiC in their dust 
shells.  They may represent an unusual population and thus 
make a poor test sample.  The trends when comparing the 
overall Galactic and SMC samples are still clear.


We have been careful to base our comparisons on sources with 
similar amounts of dust, which is quite different from a 
comparison of the amount of dust produced by similar sources
in the SMC and the Galaxy.  A definitive answer will require
radiative transfer modelling to properly measure the amount
of amorphous carbon dust and a careful assessment of the
luminosities and pulsational properties of the stars.  We
leave that work to the future, but we note in the meantime
that the IRS spectra and the derived [6.4]$-$[9.3] colors
clearly indicate that carbon stars in the SMC are capable of 
producing significant quantities of dust.



\section{Conclusions}  

Analyzing the strength of identifiable spectral features as 
a function of [6.4]$-$[9.3] color allows us to separate the
sample on the basis of the optical depth through the dust 
shell, which is effectively a measure of the amount of 
amorphous carbon around the star.  Thus we can compare the 
carbon stars in the SMC sample to sources with a similar 
amount of circumstellar dust in the Galactic sample.

Carbon stars in the SMC have lower abundances of SiC and MgS 
dust compared to the amorphous carbon dust than their 
Galactic counterparts.  This result demonstrates that the nature 
of the dust produced by stars on the AGB can vary measurably 
with their initial metallicity.  

Our observations of molecular band strengths confirm the 
shorter-wavelength results of \cite{vl99} and 
\citet[][ 2005]{mat02} who showed that stars with lower 
metallicities actually produce deeper acetylene absorption bands.  
They attributed this unexpected result to significantly higher 
C/O ratios in these stars.  We have made our comparison on 
the basis of the [6.4]$-$[9.3] color, allowing us to conclude 
that stars surrounded by similar amounts of dust have more 
acetylene around them in the SMC than in the Galaxy, and have 
proposed an alternative scenario.  The lack of SiC dust in 
the shells may deprive the amorphous carbon of seeds, lead to 
less efficient dust formation, and result in a higher ratio
of acetylene to amorphous carbon.

Finally, we note that the combination of our results and
previous work suggest that more carbon-rich dust should be
injected into the ISM in a more metal-poor galaxy.
First, a higher fraction of the stars evolve into carbon 
stars.  Second, individual carbon stars in these metal-poor
systems produce quantities of dust comparable to their
Galactic counterparts, as our spectroscopy has shown.  
Finally, oxygen-rich stars should not produce as much dust, 
since the limiting factors will be the abundances of heavier 
elements like Mg, Al, Si, and Fe, all of which will be 
reduced in low-metallicity environments.
Since amorphous carbon is the likely building block for
PAH molecules, this result has important implications for
the study of the emission features produced by these
molecules in distant and/or metal-poor galaxies.


\acknowledgments

We gratefully acknowledge the referee, J.~van~Loon, whose
thorough commentary has helped us to substantially
improve this paper, and A.~A.~Zijlstra, whose input also
proved most valuable.
Support for G.~C.~S. was provided by NASA through Contract
Number 1257184 issued by the Jet Propulsion Laboratory,
California Institute of Technology under NASA contract 1407. 
M.~M. is supported by the PPARC.  The Australian Research 
Council provided support to P.~R.~W.
This research has made use of the SIMBAD and VIZIER databases,
operated at the Centre de Donn\'{e}es astronomiques de
Strasbourg, and the Infrared Science Archive at the Infrared
Processing and Analysis Center, which is operated by JPL.



\begin{thebibliography}{}
\bibitem[Alcock et al.(1996)]{macho} Alcock, C., et al. 1996, \apj,
  461, 84
\bibitem[Allamandola et al.(1989)]{atb89} Allamandola, L.~J., 
  Tielens, A.~G.~G.~M., \& Barker, J.~R. 1989, \apjs, 71, 733
\bibitem[Aoki et al.(1998)]{ato98} Aoki, W., Tsuji, T., \& Ohnaka, K.
  1998, \aap, 340, 222
\bibitem[Aoki et al.(1999)]{ato99} Aoki, W., Tsuji, T., \& Ohnaka, K.
  1999, \aap, 350, 945
\bibitem[Blanco et al.(1978)]{bbm78} Blanco, B.~M., Blanco, V.~M., \&
  McCarthy, M.~F. 1978, Nature, 271, 638
\bibitem[Blanco et al.(1980)]{bbm80} Blanco, B.~M., Blanco, V.~M., \&
  McCarthy, M.~F. 1980, \apj, 242, 938
\bibitem[Cadwell et al.(1994)]{cad94} Cadwell, B.~J., Wang, H., Feigelson,
  E.~D., \& Frenklach, M. 1994, \apj, 429, 285
\bibitem[Cherchneff(1992)]{ch92} Cherchneff, I., Barker, J.~R., \&
  Tielens, A.~G.~G.~M. 1992, \apj, 401, 269
\bibitem[Cl\'{e}ment et al.(2003)]{cle03} Cl\'{e}ment, D., Mutschke,
  H., Klein, R., \& Henning, Th. 2003, \apj, 594, 642
\bibitem[Dufton et al.(2005)]{du05} Dufton, P.~L., Ryans, R.~S.~L., 
  Trundle, C., Lennon, D.~J., Hubeny, L., Lanz, T., \& Allende Prieto, 
  C. 2005, \aap, 434, 1125
\bibitem[Egan \& Price(1996)]{msx96} Egan, M.~P., \& Price, S.~D., 1996,
  \aj, 112, 2862
\bibitem[Egan et al.(2001)]{msx01} Egan, M.~P., Van Dyk, S.~D., \&
  Price, S.~D. 2001, \aj, 122, 1844
\bibitem[Egan et al.(2003)]{msx03} Egan, M. P., et al. 2003,
  ``The Midcourse Space Experiment Point Source Catalog, Version 2.3,
  Explanatory Guide,'' Air Force Research Laboratory Technical Report
  AFRL-VS-TR-2003-1589
\bibitem[Forrest et al.(1981)]{fhm81} Forrest, W.~J., Houck, J.~R.,
  \& McCarthy, J.~F. 1981, \apj, 248, 195
\bibitem[Frenklach et al.(1989)]{fr89} Frenklach, M., Carmer, C.~S.,
  \& Feigelson, E.~D. 1989, Nature, 339, 196
\bibitem[Frenklach \& Feigelson(1989)]{ff89} Frenklach, M. \&
  Feigelson, E.~D. 1989, \apj, 341, 372
\bibitem[Gehrz(1989)]{geh89} Gehrz, R.~D., 1989, in Interstellar Dust, 
  Proc. IAU Symp. 135, ed. L.~J. Allamandola \& A.~G.~G.~M. Tielens 
  (Dordrecht: Kluwer), 445
\bibitem[Gilman(1969)]{gil69} Gilman, R.~C. 1969, \apj, 155, L185
\bibitem[Goebel \& Moseley(1985)]{gm85} Goebel, J.~H. \& Moseley, S.~H.
  1985, \apj, 290, L35
\bibitem[Griffin(1990)]{gri90} Griffin, I.~P. 1990, \mnras, 247, 591
\bibitem[Hackwell(1972)]{hac72} Hackwell, J.~A. 1972, \aap, 21, 239
\bibitem[Hill et al.(1997)]{hi97} Hill, V., Barbuy, B., \& Spite, M.
  1997, \aap, 323, 461
\bibitem[Hony et al.(2002)]{hwt02} Hony, S., Waters, L.~B.~F.~M., \&
  Tielens, A.~G.~G.~M. 2002, \aap, 390, 533
\bibitem[Houck et al.(2004)]{hou04} Houck, J.~R., et al. 2004, \apjs, 154, 18
\bibitem[{\it IRAS} Science Team(1986)]{lrs86} {\it IRAS} Science Team 1986,
  \aaps, 65, 607 (LRS Atlas)
\bibitem[Jones et al.(1978)]{jon78} Jones, B., Merrill, K.~M., Puetter, 
  R.~C., \& Willner, S.~P. 1978, \aj, 83, 1437
\bibitem[J{\o}rgensen et al.(2000)]{jhl00} J{\o}rgensen, U.~G., Hron, J.,
  \& Loidl, R. 2000, \aap, 356, 253
\bibitem[Kova\v{c}avi\'{c} et al.(2005)]{kov05} Kova\v{c}avi\'{c}, E.,
  Stefanovi\'{c}, I., Berndt, J., \& Winter, J. 2003, J. Appl. Phys., 93, 2924
\bibitem[Kova\v{c}avi\'{c} et al.(2003)]{kov03} Kova\v{c}avi\'{c}, E.,
  Stefanovi\'{c}, I., Berndt, J., Pendleton, Y.~J., \& Winter, J. 
  2005, \apj, 623, 242
\bibitem[Kraemer et al.(2002)]{kra02} Kraemer, K. E., Sloan, G. C.,
  Price, S. D., \& Walker, H. J. 2002, \apjs, 140, 389
\bibitem[Kraemer et al.(2005)]{kra05} Kraemer, K.~E., Sloan, G.~C.,
  Wood, P.~R., Price, S.~D., \& Egan, M.~P. 2005, \apjl, 631, L147
\bibitem[Kurt et al.(1999)]{ku99} Kurt, C.~M., Dufour, R.~J., Garnett,
  D.~R., Skillman, E.~D., Mathis, J.~R., Peimbert, M., Torres-Peimbert,
  S., \& Ruiz, M.-T. 1999, \apj, 518, 246
\bibitem[Lattanzio \& Wood(2004)]{lw04} Lattanzio, J.~C. \& Wood, P.~R. 2004, 
  ``Evolution, Nucleosynthesis and Pulsation of AGB Stars,'' in Asymptotic 
  Giant Branch Stars, ed. H.~J. Habing \& H. Olofsson, A\&A Library
  (New York:  Springer), 23
\bibitem[Lennon et al.(2003)]{le03} Lennon, D.~J., Dufton, P.~L., \& 
  Crowley, C. 2003, \aap, 398, 455
\bibitem[Luck et al.(1998)]{lu98} Luck, R.~E., Moffett, T.~J., Barnes,
  T.~G., III, \& Gieren, W.~P. 1998, \aj, 115, 605
\bibitem[McGregor(1994)]{mcg94} McGregor, P.~J. 1994, PASP, 106, 508
\bibitem[McGregor et al.(1994)]{caspir} McGregor, P.~J., Hart, J.,
  Hoadley, D., \& Bloxham, G. 1994, in Infrared Astronomy with Arrays, 
  ed. I McLean (Dordrecht:  Kluwer), 299 
\bibitem[Martin \& Rogers(1987)]{mr87} Martin P.~G. \& Rogers C. 1987,
  \apj, 322, 374
\bibitem[Matsuura et al.(2002)]{mat02} Matsuura, M., Zijlstra, A.~A., van 
  Loon, J.~Th., Yamamura, I., Markwick, A.~J., Woods, P.~M., \& Waters, 
  L.~B.~F.~M. 2002, \apj, 580, L133
\bibitem[Matsuura et al.(2005)]{mat05} Matsuura, M., Zijlstra, A.~A., van
  Loon, J.~Th., Yamamura, I., Markwick, A.~J., Whitelock, P.~A., Woods, P.~M.,
  Marshall, J.R., Feast, M.~W. \& Waters, L.~B.~F.~M. 2005, \aap, 434, 706
\bibitem[Matsuura et al.(2006)]{mat06} Matsuura, M., et al. 2006, \mnras,
  submitted
\bibitem[Olivier \& Wood(2005)]{ow05} Olivier, E.~A. \&  Wood, P.~R. 2005, 
  \mnras, 362, 1396
\bibitem[Press et al.(1988)]{pr88} Press, W. H., Flannery, B. P.,
  Teukolsky, S. A., \& Vetterling, W. T. 1988, Numerical Recipes in C
  (Cambridge:  Cambridge Univ. Press)
\bibitem[Renzini \& Voli(1981)]{rv81} Renzini, A. \& Voli, M. 1981, \aap,
  94, 175
\bibitem[Russell \& Bessell(1989)]{rb89} Russell, S.~G. \& Bessell, M.~S.
 1989, \apjs, 70, 865
\bibitem[Skrutskie et al.(2006)]{2mass} Skrutskie, M.~F., et al. 2006, 
  \aj, 131, 1163
\bibitem[Sloan et al.(2005)]{sl05} Sloan, G.~C., Spoon, H.~W.~W.,
  \& Bernard-Salas, J. 2005, IRS Technical Report 05002,
  Low-resolution wavelength calibration of the IRS (Ithaca, NY:
  Cornell, available at http://isc.astro.cornell.edu/tech/tr/)
\bibitem[Sloan \& Egan(1995)]{se95} Sloan G.~C. \& Egan M.~P. 1995, \apj, 444, 452
\bibitem[Smith et al.(2002)]{smi02} Smith, V.~V., et al. 2002, \aj, 124, 3241 
\bibitem[Speck et al.(1997)]{spe97} Speck, A.~K., Barlow, M.~J. \& Skinner,
  C.~J. 1997, \mnras, 288, 431
\bibitem[Speck et al.(2005)]{spe05} Speck, A.~K., Thompson, G.~D., \&
  Hofmeister, A.~M. 2005, \apj, 634, 426
\bibitem[Szym\'{a}nski(2005)]{ogle05} Szyma\'{n}ski, M.~K. 2005, Acta Astron., 
  55, 43
\bibitem[Treffers \& Cohen(1974)]{tc74} Treffers, R. \& Cohen, M. 1974, \apj,
  188, 545
\bibitem[Trundle et al.(2004)]{tr04} Trundle, C., Lennon, D.~J., Puls,
  J., Dufton, P.~L. 2004, \aap, 417, 217
\bibitem[Udalski et al.(1997)]{ogle97} Udalski, A., Kubiak, M., \&
  Szym\'{a}nski, M. 1997, Acta Astron., 47, 319
\bibitem[van Loon et al.(1997)]{vl97} van Loon, J.~Th., Zijlstra, A.~A.,
  Whitelock, P.~A., Waters, L.~B.~F.~M., Loup, C., \& Trams, N.~R. 1997
  \aap, 325, 585
\bibitem[van Loon et al.(1999)]{vl99} van Loon, J.~Th., Zijlstra, A.~A.,
  \& Groenewegen, M.~A.~T. 1999, \aap, 346, 805
\bibitem[van Loon et al.(2006)]{vl06} van Loon, J.~Th., Marshall, J.~R., 
  Cohen, M., Matsuura, M., Wood, P.~R., Yamamura, I., Zijlstra, A.~A. 2006,
  \aap, 447, 971
\bibitem[Vassiliadis \& Wood(1993)]{vw93} Vassiliadis, E. \& Wood P.~R. 1993, 
  \apj, 413, 641
\bibitem[Werner et al.(2004)]{wer04} Werner, M.~W., et al. 2004,
  \apjs, 154, 1
\bibitem[Wheeler et al.(1989)]{wst89} Wheeler, J.~C., Sneden, C., \&
  Truran, J.~W., Jr. 1989, \araa, 27, 279
\bibitem[Whitelock et al.(2003)]{whi03} Whitelock, P.~A., Feast, M.~W.,
  van Loon, J.~Th., \& Zijlstra, A.~A. 2003, \mnras, 342, 86
\bibitem[Wood(1981)]{woo81} Wood, P.~R. 1981, in Physical Processes in
  Red Giants, ed. I. Iben \& A. Renzini, (Dordrecht:  Reidel), 135
\bibitem[Wood(1998)]{woo98} Wood, P.~R. 2998, \aap, 338, 592
\bibitem[Wood(2003)]{woo03} Wood, P.~R. 2003, in Mass-Losing Pulsating Stars 
  and their Circumstellar Matter, ed. Y. Nakada, M. Honma \& M. Seki, 
  (Dordrecht:  Kluwer), 3
\bibitem[Wood et al.(1992)]{woo92} Wood, P.~R., Whiteoak, J.~B.,
  Hughes, S.~M.~G., Bessell, M.~S., Gardner, F.~F., \& Hyland, A.~R.
  1992, \apj, 397, 552
\bibitem[Yang et al.(2004)]{ych04} Yang, X., Chen, P., \& He, J. 2004,
  \aap, 414, 1049
\bibitem[Zijlstra et al.(2006)]{mcpw06} Zijlstra, A.~A., et al. 2006, \mnras,
  submitted
\end{thebibliography}
\end{document}